\documentclass{article}
\usepackage{natbib}
\usepackage[left=1.3in,top=1in,right=1.3in,bottom=1.4in,nohead]{geometry}
\usepackage{setspace}
\usepackage{appendix}
\usepackage[ngerman,english]{babel}
\usepackage{amsmath}
\usepackage{theorem}
\usepackage{amsfonts}
\usepackage{amssymb}
\usepackage{amsbsy}
\usepackage{fancyhdr}
\usepackage{booktabs}
\usepackage{verbatim}
\usepackage{graphicx}
\usepackage{subfigure}
\usepackage{booktabs,caption}
\usepackage{longtable} 
\usepackage{color,soul}
\usepackage[shortlabels]{enumitem}
\usepackage{longtable}
\usepackage{setspace}
\usepackage{threeparttable}
\usepackage{ulem}
\usepackage{authblk}
\usepackage{url}
\usepackage{natbib}
\usepackage{url}
\usepackage{hyperref}
\usepackage{quotes}

\graphicspath{ {./graphs/} }
\title{'Moving On' - Investigating Inventors' Ethnic Origins Using Supervised Learning}

\author{Matthias Niggli\thanks{Correspondence address: Matthias Niggli, Center of International Economics and Business CIEB, University of Basel, Peter Merian-Weg 6, CH-4002 Basel, Switzerland, email: matthias.niggli@unibas.ch. I am extremely grateful for numerous helpful discussion with Christian Rutzer. I also thank Kerstin Hansen, Aya Kachi, Christoph Thommen, Sinan Acemoglu, Thomas Gerber, Rolf Weder, Dragan Filimonovic and Gabriele Cristelli for comments, inputs and suggestions. Furthermore, I am thankful to Junting Ye and Steven Skiena for granting me access to the NamePrism API on multiple occasions. This paper is an adapted version of an essay that is part of the author's dissertation. All remaining errors are my own.}}
\affil{University of Basel, Switzerland}
\date{\today}

\begin{document}

\onehalfspacing

\maketitle

\begin{abstract}
Patent data provides rich information about technical inventions, but does not disclose the ethnic origin of inventors. In this paper, I use supervised learning techniques to infer this information. To do so, I construct a dataset of $95'202$ labeled names and train an artificial recurrent neural network with long-short-term memory (LSTM) to predict ethnic origins based on names. The trained network achieves an overall performance of $91$\% across $17$ ethnic origins. I use this model to classify and investigate the ethnic origins of $2.68$ million inventors and provide novel descriptive evidence regarding their ethnic origin composition over time and across countries and technological fields. The global ethnic origin composition has become more diverse over the last decades, which was mostly due to a relative increase of Asian origin inventors. Furthermore, the prevalence of foreign-origin inventors is especially high in the USA, but has also increased in other high-income economies. This increase was mainly driven by an inflow of non-western inventors into emerging high-technology fields for the USA, but not for other high-income countries.
\end{abstract}

\newpage

\section{Introduction}

In a 2019 report, McKinsey, a consulting company, asked around $100$ CEOs about the most important factors for their business location decisions. The clear winner in this survey was talent availability \cite[]{McKinsey2019}. This important role of access to skilled workers has long been reflected in the economics literature. In recent years, especially the role of migration and ethnic diversity for innovation, development and economic growth has gained increasing attention.\footnote{See, for example, \cite{bloom2019, lissoni2018, miguelez2018, kemeny2017, Kerr2016, Peri2016} for some recent contributions.} In particular, researchers often use data on patent inventors to investigate the relationship between migration flows and inventive activity.\footnote{Prominent contributions include, for example, \cite{KerrKerr2018, miguelez2018, Breschi2017, MiguelezFink2013, kerrlincoln2010} or \cite{kerr2007}.} However, there is an important challenge associated with this field of research: Inventors' nationalities are generally not observed on patents and their ethnic origins have to be inferred based on their names. Typically, existing approaches from the economics literature rely on commercial databases for this endeavour and use name-matching techniques to assign patent inventors to ethnic origins.

This is where this paper differs from previous approaches and contributes to the literature. In particular, I propose a new approach to determine patent inventors' ethnic origins that is based on supervised learning techniques and addresses several challenges of preceding methods. To do so, I construct a dataset of $95'202$ labeled names and use it to train an artificial neural network that directly classifies patent inventors' ethnic origins based on their names. Building on this classification approach, I then present novel insights regarding the ethnic origin composition of inventors across countries and technological fields over a period of $35$ years. To the best of my knowledge, this paper is the first to propose a procedure for patent inventor classification that is based on supervised learning. Using such an approach allows me to tackle several challenges associated with preceding methods. First, it can better address the high precision/low recall problem that is inherent to more conventional name matching techniques.\footnote{See Section \ref{classification_sec} for more inforamtion and, for example, \cite{karaulova2019} and \cite{Breschi2017} for a extensive discussion on this issue.} Second, while many preceding studies had to focus on a rather small set of ethnic origins, my proposed approach allows to extend inventor classification to $17$ ethnic origin groups. Third, the approach does not rely on a particular host country and can be used to investigate the ethnic origin composition of inventors internationally. Fourth, it mitigates mis-classification problems arising from spelling errors in inventor names. Fifth, it does not require a commercial database, which makes it more accessible, transparent and allows to empirically evaluate classification performances. 

My approach builds on a publicly available dataset of athletes who participated in Olympic games between $1896$ and $2016$. To mitigate problems of unbalanced data (i.e. athletes of some ethnic origins are more frequent in this sample than others), I construct an additional dataset of labeled names using information from NamePrism \cite[]{Ye2017}. Combining both datasets results in my final training sample of $95'202$ labeled names. I then use this dataset to train a classification model that learns to predict inventors' ethnic origins based on their names. The model I use is an artificial recurrent neural network with long short-term memory (LSTM). This neural network uses the letters a name consists of as input data and outputs probability scores for belonging to $17$ different ethnic origins. After training and optimization, the model achieves an overall classification performance of $91.0$\% across these $17$ ethnic origins when evaluated on a testing set (based on a weighted F1-score). This is a relatively high performance with regard to related approaches in the literature that report performance scores \cite[see e.g.][]{karaulova2019,Ye2017}. Using this trained classification model, I predict and investigate the ethnic origins of over $2.68$ million inventors from patents filed at the European Patent Office (EPO) or the United States Patent and Trademark Office (USPTO) between $1980$ and $2015$. The global ethnic origin composition of inventors has become more diverse over this time period, which was mostly due to an increase of inventors with Asian backgrounds. Furthermore, I find a substantially more diverse inventor composition in the USA compared to other high-income countries. In particular, the USA has witnessed large inflows of non-western origin inventors, particularly to emerging high-technology fields. A similar trend is missing in most European countries.

This paper is most closely related to the seminal contribution from \cite{kerr2007}, who also determines inventors' ethnic origins. \cite{kerr2007} uses a commercial database and a name matching procedure to classify inventors to nine ethnic groups and studies their ethnic composition for the case of the USA. My paper allows to refine and extend this analysis in two particular ways. First, I use supervised learning techniques to classify inventors to $17$ ethnic origin groups, which also enables me to evaluate the statistical performance of my approach. Second, this more granular classification taxonomy allows me to extend \cite{kerr2007}'s seminal analysis of the ethnic origin composition of inventors beyond the case of the USA. With this latter regard, my paper is also related to \cite{miguelez2018} and \cite{MiguelezFink2013}, who both study inventor compositions across countries. However, different to my approach, these authors rely on a rather narrow and specific datatset of patents and focus on the nationality of inventors and not on their ethnic origins.
Besides the contributions from the above-mentioned authors, there are further papers in the literature that propose to use names for inventor origin classification. My paper primarily differs from these contributions with regard to their methodological approaches as well as the scopes of their analyses, which are often restricted to one specific country or ethnic origin. For example, \cite{karaulova2019} present a stepwise procedure that is primarily based on the morphology
of surnames. Their proposed approach does not rely on a commercial database, but they focus exclusively on Russian origin. \cite{ferrucci2020} and \cite{Breschi2017} use similar name matching approaches as \cite{kerr2007} that are also based on commercial databases.\footnote{\cite{ferrucci2020} uses "Forebears" and \cite{Breschi2017} the "IBM Global Name Recognition" database.} While \cite{ferrucci2020} focuses specifically on Soviet inventors and investigates their contribution to German patenting after the Cold War, \cite{Breschi2017} study patent collaborations between domestic and foreign origin inventors in the USA. 
Lastly, my paper is also broadly related to several studies that focus more generally on ethnic origin classification but do not have a particular focus on inventors.\footnote{For example, \cite{Ye2017} construct so-called "name embeddings" and classify names to a hierarchical set of 39 ethnic groups, \cite{treeratpituk2012} assign names to 12 origins using multinomial logistic regressions and \cite{ambekar2009} choose hidden markov models and decision trees to map names to 13 ethnic groups.}

The remainder of this paper is structured as follows. In Section \ref{classification_sec}, I review existing approaches to classify inventors' origins and introduce the one used in this paper. I present the dataset used for training my classification model, document its training procedure and report the model's classification performance. In Section \ref{ethnic_sec}, I use the trained model to predict the ethnic origins of inventors from a large sample of patents. I provide descriptive evidence about the ethnic origin composition across countries, technological fields and over time, and discuss notable differences between the USA and other high-income economies. Finally, Section \ref{conclusion} discusses the overall findings and concludes the paper.

\section{Classification of Inventors' Ethnic Origins}\label{classification_sec}

Patent data is widely used and acknowledged for economic research. However, most patents do not provide information about the nationality or ethnicity of its inventors. This makes it difficult to analyze migration trends in patent statistics. Researchers have thus developed approaches to \textit{infer} migrant backgrounds of inventors. The common idea of these approaches is to focus on inventors' ethnic origins. Unlike nationalities, ethnic origins can be derived from inventors' names which are always stated on patent documents \cite[see e.g.][]{KerrKerr2018, Breschi2017, Agrawal2008, kerr2007}. From an empirical point of view, the main question is then how to relate names to ethnic origins. Pioneer work by \cite{kerr2007} has used "Melissa", a commercial database, which contains frequent names for different ethnic origins. Inventors' names were matched to these names in the database and if a match could be found, the inventor was assigned to the corresponding country of origin stated in the "Melissa" database. This approach has proved to be useful and has been adopted extensively in the economics literature \cite[e.g.][]{ferrucci2020,AkcigitStantcheva2016, ghani2014, gaule2013}. But there are some important challenges related to such name-matching techniques (see \citealt{karaulova2019} or \citealt{Breschi2017} for an overview). First of all, name matching approaches generally have a low recall, which means that they tend to miss a substantial fraction of inventors from a given ethnic origin \cite[]{karaulova2019}. This is because the reference databases that name matching approaches rely on generally feature the most popular names for a given ethnic origin, but never have full coverage for all names. Accordingly, inventors with less popular names cannot be matched and become false negatives. Another challenge is the ambiguity of names in patent data. For example, simple spelling errors like "Francesco" (an Italian name) instead of "Francisco" (a Hispanic name) can introduce bias \cite[]{Breschi2017}. A further potential problem could be that inventors are assigned to only one specific ethnic origin, which can be insufficient to account for migrant backgrounds. For example, an inventor named "Joaquin Smith" could be considered of Anglo-Saxon origin because of his surname even though his first name implies a Hispanic origin as well.   

An alternative for studying the composition of inventors compared to name-matching approaches is to examine a unique set of patents that were filed at the USPTO through the Patent Cooperation Treaty (PCT) system \cite[]{MiguelezFink2013}. Unlike any other set of patents, these patents state the inventors' resident addresses as well as their citizenship. The data has been made available by the World Intellectual Property Organization (WIPO) and has been extensively used by researches ever since. \cite{MiguelezFink2013} describe the data in detail and provide initial estimates of migration flows and inventor diasporas over time. Naturally, this dataset also comes with a number of limitations. For example, data coverage varies across countries (e.g. inventors from the USA, Canada or the Netherlands are underrepresented) and focusing on citizenship can bias migration estimates. This is because inventors can obtain the citizenship of their host countries and, consequently, will not be counted as migrants anymore. As regulations for obtaining citizenship differ across countries, this can introduce additional bias to estimated migration flows \cite[]{MiguelezFink2013, Breschi2017}. Most importantly, however, the data covers only a limited time window up to the year $2013$. After this date, regulatory requirements have changed and disclosing nationalities was no longer required \cite[]{MiguelezFink2013}. Therefore, the data cannot be used for analysis after $2013$ and different approaches are needed for studying later periods.
\cite{Breschi2017} have developed a refined name-matching method, which does not share these shortcomings and, additionally, overcomes some of the challenges of preceding approaches. First, these authors use a specific database to obtain a set of unique inventors from patents filed at the EPO. Subsequently, they retrieve frequencies of these inventor names within and across so-called "Countries of Association" listed in the IBM Global Name Recognition database (IBM-GNR). This information can be used to construct indicators for a name belonging to any potential ethnic origin.\footnote{The advantage of using the so-called "EP-INV" inventor database is that names of inventors are controlled for spelling errors. In their subsequent analysis, \cite{Breschi2017} focus on inventor migration to the USA. Consequently, they define indicators according to their specific needs, i.e. countries of origin are assigned to inventors according to three specific indicators, one of which controls for Anglo-Saxon or Hispanic origin. In principle, different indicators and classification procedures could be used. To find optimal threshold values for migrant status assignment, \cite{Breschi2017} compare classification results from a wide range of different indicator thresholds values against a baseline sample retrieved from the above-mentioned WIPO dataset. Appendix 2 of their paper documents the technical procedure of this classification approach in great detail.} However, this approach also focuses on the matching of names in different datasets and builds on a commercial database.

This is where my approach differs from the existing economics literature. Instead of relying on name-matching techniques, I use supervised learning to relate names to ethnic origins.\footnote{Machine learning techniques (supervised, unsupervised and combinations of both) have been frequently used in the computer science literature to classify the ethnicity of names. Some examples and applications include \cite{YeSkiena2019,Ye2017,torvik2016,treeratpituk2012,chang2010,ambekar2009}.} I exploit that inventor names can be encoded as sets of letters, which can be processed by supervised learning algorithms. If these sets have patterns that differ across ethnic origins, the letters in an inventor's name can be powerful predictors for his or her ethnic origin \cite[]{webster2004}. Formally, I use the letters a name consists of as input variables $X$ for a supervised learning algorithm that relates inventors to ethnic origins $G_k$. More specifically, such an algorithm estimates the conditional class probability $Pr(G_k|X=x)$ of $k$ different ethnic origins \cite[]{FriedmanHastie2001}.

Besides not relying on matching names to their counterparts in a commercial database, this approach has some additional strengths. A first advantage refers to name-matching techniques' natural problems if coverage for some names is low in the underlying ethnic origin database. Ethnic origin assignment is obviously impossible if an inventor's name is not covered at all by the database. This is different with supervised learning methods. Instead of looking for perfect matches, these algorithms search for patterns in names and classify inventor's ethnic origins accordingly. With this regard, coverage is not an issue for my proposed method. Hence, it is well suited to approach the challenge of high precision/low recall that is inherent to name matching techniques \cite[]{karaulova2019, Breschi2017}. Furthermore, if only the inventor's first name or surname can be found, name-matching approaches are still feasible but likely become more error-prone. This is because ethnic origin assignment will be based on relatively little information in such cases. For example, matching could be exclusively based on first names, which may not be very origin-specific and can thus lead to classification errors. Consider an inventor named "Laura" whose surname cannot be found in the underlying ethnic origin database. As a consequence, her ethnic origin would be classified solely according to her first name, which is common among several ethnic origins (e.g. Hispanic, French, German, Italian, Anglo-Saxon). This does not mean that assignment will necessarily be wrong (e.g. if origin classes are rather broad), but classification becomes much more difficult and less specific because of less available information. Again, this shortcoming does not exist for supervised learning techniques. In addition, as supervised learning algorithms focus on patterns in names, problems of spelling errors are automatically mitigated. Finally, supervised learning models' performances can be transparently evaluated using established statistical methods.

The main challenge for my proposed approach is to find a model that learns a representation of the conditional class probability, $Pr(G_k|X=x)$, which minimizes mis-classification error. In order to train and validate such a model, the first step is to find a set of names that are labeled to ethnic origins. In the following subsection, I highlight how I approach this challenge and describe the data that I am using for training my classifier in detail.

\subsection*{Dataset}

Oftentimes, the preceding literature assumes that ethnic origins can be approximated well by nationalities. Given that this is a good assumption, different datasets are used. For example, \cite{ambekar2009} gather a training set consisting of celebrities listed on Wikipedia, \cite{Ye2017} construct a dataset containing celebrities and their followers on Twitter, and \cite{Breschi2017} use inventors' nationalities stated in the WIPO patent inventor database described in \cite{MiguelezFink2013}. 
In this paper, I build on a dataset consisting of athletes who participated in Olympic games from Athens 1896 to Rio 2016. The main advantage of this dataset is that it allows to use names from time periods prior to relatively large migration flows to western countries during the $20$th century. The dataset has been constructed in a tremendous effort by a group of dedicated Olympic historians and statisticians and is available on Kaggle.\footnote{For information about the data, see \url{https://www.olympedia.org/} and \url{https://olympstats.com/}. The data can be easily accessed thanks to Randi H. Griffin at \url{https://www.kaggle.com/heesoo37/120-years-of-olympic-history-athletes-and-results} (last accessed: June 8, 2021)} It features $135'571$ athletes who have been starters for $191$ different national teams and states their full names. 

In a first step, I define a selection of ethnic backgrounds which are subsequently used to investigate the ethnic origin composition of inventors across countries and over time. Two different aspects are imperative for this selection: First, it should reflect well the most prevalent inventor nationalities and ethnic groups as discussed in \cite{MiguelezFink2013} and \cite{kerr2007}. This suggests, for example, that German, Anglo-Saxon, Japanese or Indian origin should be considered. Second, it should also approximate the overall population composition in the largest patenting countries to cover significant diasporas in these countries. For example, Germany is one of the largest patenting countries and has a substantial diaspora of Turkish migrants. This suggests that Turkish origin should be considered in the selected ethnic origins, even if Turkey is not among the largest patenting countries \cite[e.g.][]{WIPO2019}.
I select $17$ different ethnic origins which I will use for analysis throughout this paper. Besides the considerations stated above, the selection of these origins is also motivated by \cite{Ye2017} who use a taxonomy that is based on cultural, ethnic and linguist (CEL) similarities proposed by \cite{mateos2007}. Again building on \cite{Ye2017}, I then assign national teams from the Olympics dataset to the selected $17$ ethnic groups, which allows me to label athletes' ethnic origins. The taxonomy of the selected $17$ ethnic origins and their corresponding national teams is presented in Table \ref{origin_list}.\footnote{National teams of the USA, Canada or Australia are not included because they have always been immigrant nations, which, as mentioned above, could distort learning. Similarly, countries such as Switzerland are excluded because their population consists of multiple ethnic origin groups (e.g. German, French and Italian for Switzerland). Ethnic origins from Sub-Saharan Africa or Central Asia are not covered because patent data does not contain a substantial number of patents and inventors from these parts of the world \cite[e.g.][]{MiguelezFink2013}. Hence, including them in the training data could also result in low(er) classification performance when ultimately applying the model to patent inventor data. Furthermore, the Philippines are not included among South-East Asian national teams because of relatively high similarities to Hispanic names \cite[]{Ye2017}.}

\begin{table}[]
\caption{Taxonomy of Ethnic Origins and National Teams}
\label{origin_list}
\centering
\begin{tabular}{ll}
\hline\hline
Ethnic Origin & National Teams / Countries \\
\hline
Anglo-Saxon & Great Britain, Ireland\\
Chinese & China\\ 
French & France\\
German & Germany\\
Hispanic-Iberian & Spain, Portugal, Mexico\\ 
India & India\\
Italian & Italy\\ 
Japanese & Japan\\ 
Korean & Korea\\ 
Arabic & Egypt, Syria, Saudi Arabia, Jordan, \\
& UAE, Tunisia, Algeria, Morocco\\
Persian & Iran\\
Slavic-Russian & Russia, Ukraine, Belarus\\
East-Europe & Poland, Czechoslovakia, Hungary\\
Balkans & Serbia, Croatia, Yugoslavia\\
Scandinavian & Sweden, Norway, Finland, Denmark, Iceland \\
South-East Asia & Vietnam, Thailand, Malaysia, Indonesia, \\
& Laos, Cambodia\\
Turkish & Turkey\\
\hline\hline
\end{tabular}
\end{table}

The chosen categories are rather broad and do not reflect detailed ethnic groups. For example, the People's Republic of China alone recognizes $56$ different ethnic groups on Chinese soil. However, building on the findings of the related literature, I believe this to be a meaningful and well-suited selection for investigating the ethnic origin composition of inventors. Similar to the related literature, the main limitation of using this dataset is rather that an athlete's national team does not always proxy ethnic backgrounds accurately. For example, in the $1996$ Olympic games in Atlanta, USA, the Chinese migrant Li Donghua won the gold medal in gymnastics for the Swiss national team. Such cases would introduce bias in my training sample and distort learning. Naturally, this problem primarily concerns athletes who were starters for national teams of countries that have experienced significant waves of immigration. From the national teams in Table \ref{origin_list}, this particularly concerns Germany, Great Britain, France, Spain, Portugal and the Scandinavian countries.\footnote{See for example \cite{dorn2021} for a recent overview of migration patterns in Europe} Already in the 1950s and 1960s some of these countries have experienced immigration of so-called "guest workers", mostly from Southern European countries and/or return-migration from former colonies. After the oil crisis in the 1970s immigration also extended to southern European countries \cite[see e.g.][]{vanMol2016}. At the same time, policies to obtain citizenship were relatively strict in many of the receiving countries at least until the $1980$s \cite[]{doomernik2016}. Thus, I expect bias to be more of a concern for athletes starting for these immigration countries in Olympic games after $1980$. Accordingly, I discard all athletes from my sample who were starters for one of the mentioned countries after the year $1980$. Because of their colonial past, I use a stricter condition for France and Great Britain. For these two countries, athletes only remain in the sample if they started in Olympic games before $1970$. This procedure restricts my sample to $42'013$ athletes from national teams stated in Table \ref{origin_list}. In principle, this could already be a reasonable overall sample size for training a classification model. But there are relatively few samples for some ethnic origins. For example, there are only $890$ athletes for India and only $793$ for Turkey. These are rather small class sizes and they might not be large enough to effectively learn name patterns for these ethnic origins. Furthermore, ethnic origin shares in my athletes dataset do most likely not always correspond well to those among patent inventors. For example, results reported by \cite{MiguelezFink2013} and \cite{kerr2007} suggest that Indian origin is important. Accordingly, some ethnic origins might not only suffer from limited sample sizes in the athletes dataset, but could also receive small weights that do not correspond to their prevalence in patent data (the reverse also being possible). In other words, athletes data is not the same as inventor data which, in turn, could bias classification results for patent inventor data. 

Such problems of unbalanced training data are a common issue in machine learning applications. In principle, they could be somewhat approached by using class-weighted loss functions in the learning process of the classification model. But, if possible, the better solution is to add more samples to the data \cite[]{Chollet2018}. To do this, I gather an additional sample of unlabeled names and infer their ethnic origins using NamePrism. NamePrism is a non-commercial ethnicity classification tool that aims to support academic research and has been used by several recent economic studies \cite[e.g.][]{deRassenfosse2020, Diamond2019}. NamePrism constructs name embeddings from a set of $74$ million labeled names from $118$ countries and uses a Naive Bayes classifier to assign names to a hierarchical set of $39$ so-called leaf nationalities \cite[]{Ye2017}.\footnote{In principle, NamePrism could be used directly to classify patent inventors' ethnic origins. However, the taxonomy that is used to train it does most likely not optimally reflect the composition of inventors \cite[see e.g.][]{MiguelezFink2013, kerr2007}. Furthermore, performance scores on some ethnic origins classes that could be important regarding inventor classification are rather small (e.g. German)} For a given name, NamePrism provides class probabilities to belong to any of the $39$ leaf nationalities. Using these class probabilities allows me to develop a procedure to assign unlabeled names to the $17$ ethnic origins stated in Table \ref{origin_list} and to construct additional training samples.

In order to do this, I sample the names of $63'585$ inventors stated on patents filed at the USPTO or the EPO.\footnote{I sampled inventor names with a resident address in the countries mentioned in Table \ref{origin_list}. Additionally, I also sampled inventor names from the USA, Canada, New Zealand, Australia, Austria, Switzerland, Argentina, Colombia, Chile, Brazil and Iraq to either reflect these countries' role in global patenting or to potentially add more names for selected ethnic origins (e.g. Iraq for Arabic or Persian origin).} I then accessed NamePrism's API and retrieved these names' probabilities to belonging to the $39$ NamePrism leaf nationalities. After collecting this information, the question is then how to best map the NamePrism leaf nationality predictions to the $17$ ethnic origin classes $G_k$ that I am using in this paper. One possibility is to manually construct a crosswalk between leaf nationalities and ethnic origin classes (an example for such a crosswalk is presented in Table \ref{NamePrismCrosswalk} in the Appendix). However, as this has to be performed manually, it could be wrong or at least overly simplified. An alternative is using supervised learning techniques. In essence, mapping leaf nationalities to ethnic origins boils down to a classification problem: The goal is to classify a name to one of the 17 ethnic origins based on the $39$ leaf nationality predictions collected from NamePrism. Typically, this is a task where supervised learning techniques perform very well. Therefore, a likely better approach compared to manually defining a crosswalk, is to train a classification algorithm that learns an optimal mapping function between the two taxonomies. To test this and learn such a function, I first gather the necessary data: I construct a stratified sample of $12'526$ labeled names from the athletes dataset and retrieve NamePrism leaf nationality predictions for them. Using this dataset, I can then train different algorithms that learn to classify these names to the $17$ ethnic origins, $G_{k}$, based on the $39$ NamePrism leaf nationality predictions. I train three well established classifiers and compare their performance on this task: A Random Forest, Gradient Boosted Trees (XGBoost) and a Feed-Forward Artificial Neural Network. The details of the procedure are provided in Appendix \ref{TrainingData}. Evaluating these algorithms and the manual approaches on a testing set shows that the former outperform the latter by around $20$ percentage points. This corresponds to a substantial performance increase of about $33\%$. XGBoost was the best performing algorithm and, accordingly, I select and use this algorithm to classify the $63'585$ sampled inventor names. In a final step, I subset these names to those, whose ethnic origin, $G_k$, is very clearly identified. This is an important step to prevent the final name classification model to learn from falsely labeled training samples. The details for this subsetting procedure are also provided in Appendix \ref{TrainingData}. Ultimately, this results in a set of $53'189$ additional training samples which I combine with the $42'013$ athletes names. Hence, my final dataset consists of $95'202$ name samples across $17$ ethnic origins. Table \ref{TrainingDataDistribution} in the Appendix provides an overview of the class frequencies in this final training dataset.

\subsection*{Data Processing, Learning and Model Performance}

Next, I have to encode these training samples to a form which can be processed by machine learning algorithms. This is straightforward for the target variable, i.e. the ethnic origin, which can simply be labeled to one of the $17$ ethnic origins. It is somewhat more challenging to encode names as features (often referred to as "tokenization"). First, inventors' names are cleaned of any kind of punctuation and, if they exist, non-Latin letters are transformed into their Latin counterparts. Second, names can have first names, middle names and last names and, naturally, differ in length. For training, this has to be harmonized because all encoded names have to be of identical shape. Therefore, I truncate all names at $30$ letters (including white spaces between first, middle and last names). As this leaves over $95$\% of names unaffected, I do not expect this to introduce bias for learning. Next, I use one-hot-encoding of these $30$ letters to transform each name to a sparse matrix of zeros and ones. This matrix is of shape $30\times28$, whereas rows indicate the 1st, 2nd… 30th letter of each name and the $28$ columns correspond to the $26$ letters of the Latin alphabet, white-space and padding.\footnote{Padding represents the end of a name; for example, the name "Mahatma Gandhi" would have padding instances from the $15$th to the $30$th row, representing that his name only consists of $13$ letters and one white space. A visual example is given in Appendix \ref{AppendixGraphs}} After these steps, the entire set of $95'202$ encoded names is represented by a tensor of shape $95'202\times30\times28$, which can be used for training.

The algorithm I train to classify ethnic origins is a recurrent artificial neural network with long short-term memory (LSTM). This class of neural networks consists of one or more hidden layers with so-called long short-term memory that have been first proposed by \cite{hochreiter1997}. LSTMs are capable to take the sequential nature of input data into account and can detect patterns in input sequences over arbitrary long lags. Accordingly, they have been extensively used to process sequential data, for example for text analysis or to predict stock prices. A detailed overview of the LSTM architecture can be found, for example, in \cite{Goodfellow2016}.\footnote{Figure \ref{lstm_cell} in Appendix \ref{AppendixGraphs} provides a graphical illustration of a LSTM cell from these authors.} Intuitively, LSTM-cells can "memorize" and "forget" past information. Regarding names, 
this means, for example, that a LSTM-cell can "remember" (or "forget") important (or unimportant) patterns from the name's first five letters when it "sees" the sixth letter. From a more technical prospective, the network's LSTM-layers each consist of a number of $c$ LSTM-cells. These cells have an input gate, a forget gate, an output gate and an internal cell state. The internal cell state is particularly important as it reflects the cell's "memory" at a given point $s$ of the sequence. Whenever new information is processed by the network (e.g. the next letter of a name), non-linear transformations of this new input, $x_s$, together with all the LSTM-cells outputs in the same LSTM-layer, $\mathbf{h}_{s-1}$, define the output gate, the input gate and the forget gate. The input gate and the forget gate then update the internal cell state jointly with $x_s$ and $\mathbf{h}_{s-1}$. Finally, the new cell state and the output gate determine the cell's next output $h_{c,s}$ \cite[see e.g.][for a more detailed overview]{Goodfellow2016}. All of these operations are defined by trainable weights that control what to remember and what to forget for a specific LSTM-cell. During the network's training process the weights are constantly updated to minimize mis-classification error. In other words, LSTM-cells can have their focus on different patterns of the input data and learn to keep useful information in memory. Accordingly, their outputs can be thought of as abstract representations of the sequential input data. This abstract representations contain predictive power for the given classification task but are not necessarily meaningful for humans. In multi-layer neural networks, they are passed forward to subsequent layers that learn additional representations. Ultimately, an output-layer uses this abstracted information to predict a name's probability to belonging to any of the $17$ ethnic origins. 

I build a network consisting of $3$ hidden LSTM-layers with $512$, $256$ and $64$ LSTM-cells, respectively. Appendix \ref{LSTMModel} provides the details about the network's architecture and the training process, in which this network learns over $1.98$ million weights. After optimization, I evaluate the network's performance on a testing set. This provides important information on how well the network is capable to classify ethnic origins. I use F1-scores as a performance metric, which are particularly useful if it is unclear whether low recall (many false negatives) or low precision (many false positives) is a worse type of error.\footnote{F1-scores state the harmonic mean between precision and recall across classes. Precision states the share of inventors from a predicted origin who are indeed of this origin. Recall, in turn, states the share of inventors form a given ethnic origin that are also predicted to be of this origin. Formally, the F1 score is calculated as $F1=2\times(\frac{Recall \times Precision}{Recall + Precision})$.} For the classification of ethnic origins, this is exactly the case and, accordingly, F1-scores are the preferred evaluation metric in the literature \cite[]{Ye2017}. 
Table \ref{performance} presents information regarding precision, recall and F1-score on the testing set that the model has never seen before. The model's overall F1-score of $0.91$ in the first row is the weighted average of the $17$ class-specific F1-scores, using their sample frequencies as weights. Class-specific F1-scores modestly differ across ethnic origins, whereas the model shows the lowest, but still reasonable performance scores for the two smallest groups "Balkans" and "South-East Asia". 

\begin{table}[!h]
\caption{Performance of the LSTM Classification Model}
\label{performance}
\centering
\begin{threeparttable}
\begin{tabular}{lllr}
\hline\hline
Ethnic Origin & Precision & Recall & F1 Score \\
\hline
\textbf{Overall} (weighted) & \textbf{0.910} & \textbf{0.910} & \textbf{0.910} \\
& \\
Anglo-Saxon & 0.859 & 0.880 & 0.873\\
Arabic & 0.911 & 0.927 & 0.919 \\
Balkans & 0.816 & 0.753 & 0.783 \\
Chinese & 0.932 & 0.938 & 0.938 \\
East-Europe & 0.908 & 0.913 & 0.910 \\
French & 0.911 & 0.893 & 0.902 \\
German & 0.805 & 0.860 & 0.832 \\
Hispanic-Iberian & 0.908 & 0.925 & 0.916 \\
India & 0.910 & 0.852 & 0.880 \\
Italian & 0.955 & 0.901 & 0.927 \\ 
Japanese & 0.972 & 0.993 & 0.982 \\ 
Korean & 0.925 & 0.955 & 0.940 \\ 
Persian & 0.897 & 0.902 & 0.900 \\
Scandinavian & 0.919 & 0.895 & 0.907 \\
Slavic-Russian & 0.961 & 0.956 & 0.958 \\
South-East Asia & 0.871 & 0.758 & 0.810 \\
Turkish & 0.923 & 0.966 & 0.944 \\
\hline\hline
\end{tabular}
\begin{tablenotes}[flushleft]
\item \footnotesize{\textit{Notes:} The table reports performance scores from the testing set. The overall scores in the first row is the weighted average across ethnic origins, using sample frequencies as weights.}
\end{tablenotes}
\end{threeparttable}
\end{table}

It is difficult to directly compare these performance results to those of existing approaches in the literature. The reason is that some related contributions cannot report performance scores \cite[e.g.][]{kerr2007}, can have a different focus \cite[e.g.][]{Breschi2017} or use different datasets and origin taxonomies \cite[e.g.][]{Ye2017, ambekar2009}. However, to put the LSTM network's performance into perspective, the overall results of some related contributions are still interesting. From the economics literature, \cite{Agrawal2008} use name-matching techniques and focus on Indian origin only. They have no information on recall, but report a precision of $97$\% based on phone calls to $2'256$ persons who they classified to have an Indian origin. With this regard, my proposed LSTM classifier achieves a comparable precision of $91$\% (F1-score: $88$\%) for Indian names while it is able to classify $16$ additional ethnic groups instead of just one. \cite{Breschi2017} identify migrant inventors in the USA by using name-matching techniques and a weighting procedure to classify inventors' names to $10$ different "countries of origin". They do not report performance scores but provide two graphical examples for recall-precision combinations regarding Italian and Chinese origin. For both cases, the LSTM network achieves a much higher performance and substantially outperforms their illustrative examples.\footnote{
\cite{Breschi2017} maximize the class-specific recall conditional on a precision of at least $0.3$ to capture relatively many potential migrants. In their two graphical examples for Italy and China, the recall never surpasses $0.75$ based on the precision condition. This is much lower compared to the LSTM in this paper that achieves F1-scores over $90\%$ for the two ethnic groups.} In contrast to these studies building on name-matching techniques, \cite{karaulova2019} use a classification procedure for Russian names based on surname morphology. They report a F1-score of 96.0\%, which is practically identical to the 95.8\% of my LSTM model for Slavic-Russian origin. However, they exclusively focus on Russian backgrounds, whereas my approach is capable to identify a total of 17 ethnic origins. \cite{Ye2017} is another contribution which does not rely on name matching. These authors report the performance of their own supervised learning approach and also compare it to those developed by \cite{ambekar2009}, \cite{treeratpituk2012} and \cite{torvik2016}. Depending on the testing set, the mentioned methods achieve overall F1-scores between $0.36$ and $0.80$ when classifying $13$ different ethnic groups, and between $0.57$ and $0.83$ for $10$ ethnic groups. In comparison to these results, LSTM again seems to perform similarly or substantially better based on the $17$ ethnic group taxonomy used in this paper.\footnote{\cite{Ye2017}'s NamePrism classifier outperforms the other mentioned approaches on most ethnic groups and datasets. Note that NamePrism is originally trained to classify $39$ different ethnic groups, which is more than double the number of classes used in this paper. Therefore, their ethnic origin taxonomy also includes ethnic groups such as "Baltics" or "Maghreb" with relatively little training samples. Despite this granular level, NamePrism achieves a high overall F1-score of $0.806$ on this taxonomy. For comparable ethnic groups (e.g. "Celtic English" or "French"), NamePrism performs slightly lower compared to the LSTM network which could, however, be due to the larger number of groups it was trained on.}

Taken together, the LSTM's overall F1-score of $0.91$ seems to be a relatively high performance when compared to related contributions in the literature. This suggests that a neural network with LSTM architecture is a powerful tool for classifying ethnic origins based on names. That said, I do not claim that my proposed supervised learning approach is strictly more appropriate for any application of (economic) analysis. Yet, it could be very well suited to tackle the high precision/low recall problem of established name-matching approaches in the economics literature, as is mentioned, for example, by \cite{Breschi2017}. This is especially the case, if researchers aim to examine questions that involve to focus on different ethnic origins at the same time. In the next section, I exploit this particular strength of my approach and use the trained LSTM model to predict the ethnic origins of patent inventors from a large-scale dataset. This allows me to investigate the ethnic origin composition of inventors in detail across countries, technological fields and over time.

\section{The Ethnic Origin Compositions of Inventors}\label{ethnic_sec}

I predict ethnic origins of $2.68$ million inventors listed on patents filed at the USPTO or the EPO since $1980$. This dataset contains inventors from all over the world and allows me to investigate the ethnic composition of inventors across countries, technological fields and over a long period of time.\footnote{For more information regarding the patent data used in this paper see Appendix \ref{PatentData}.)} To derive and investigate these compositions, there are two possibilities: One is to assign each inventor to one specific ethnic origin according to the highest predicted probability from the LSTM classification model. Another approach is to aggregate ethnic origin probabilities from the inventor-level to a more aggregate level (e.g. the country-level) and divide it by the total number of inventors. This second approach provides a measure of the overall prevalence of different ethnic origins for different levels of aggregation. I choose to use the latter approach because it better accounts for names indicating multiple ethnic backgrounds.\footnote{Recall the previously mentioned hypothetical inventor "Joaquin Smith". Instead of assigning this person to just one ethnic origin, he could be e.g. considered $50$\% Anglo-Saxon and $50$\% Hispanic.} For a given level of aggregation $j$ (e.g. a country, technological field or region) and year $t$, I sum up all inventors' class probabilities for belonging to ethnic origin $k$ and divide this figure by the total number of registered inventors $N_{j,t}$ in this year. Formally:

\vspace{-0.4cm}
\begin{equation}
\hat{\pi}^{k}_{j,t} = \frac{\sum_{i=1}^{N_{j,t}}\hat{G}_{i,k,j,t}}{N_{j,t}},
\vspace{-0.1cm}
\end{equation}

where $\hat{G}_{i,k,j,t}$ is the predicted probability of inventor $i$ in country/technology field/region $j$ on a patent filed in year $t$ to belonging to ethnic origin $k$. All the $17$ resulting ethnic origin prevalence indicators, $\hat{\pi}^{k}_{j,t}$, sum up to unity for a given year $t$ and aggregation level $j$. 

\begin{figure}[ht]
    \caption{The Global Ethnic Origin Composition of Inventors (1980-2015)}
    \centering
    \includegraphics[scale=0.6]{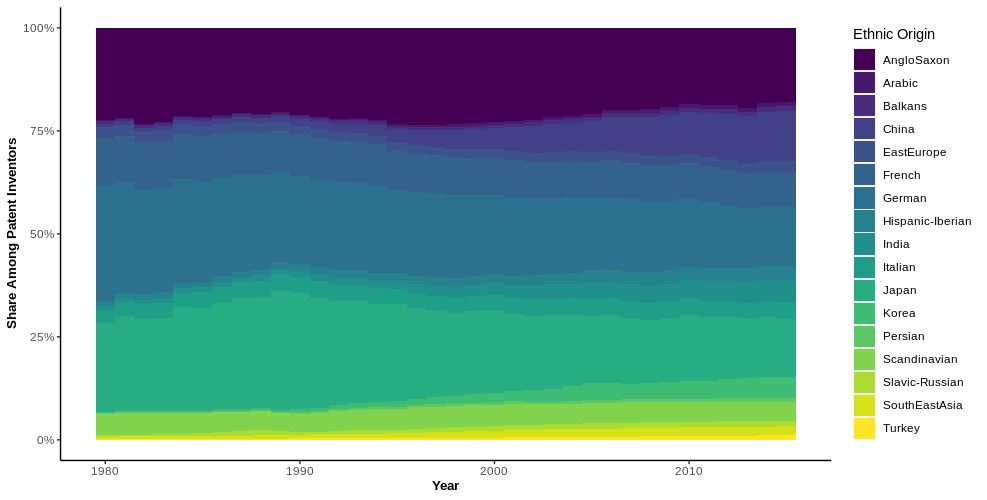}
    \captionsetup{textfont=it}
	\captionsetup{singlelinecheck = false, format= hang, justification=raggedright, font=footnotesize, labelsep=space}
	\caption*{Notes: The graph shows the evolution of the prevalence of $17$ ethnic origins among the annual stock of patent inventors between 1980 and 2015. Formally, it plots $\hat{\pi}^{k}_{j,t}$. The data for this plot is from the OECD and the USPTO.}
    \label{origin_dist}
\end{figure}

As a starting point, Figure \ref{origin_dist} plots the distribution of $\hat{\pi}^{k}_{j,t}$ at the global level. Anglo-Saxon, German and Japanese origin are the most prevalent ethnic origins across time. However, all of their shares have decreased since the $1990$s. Interestingly, the prevalence of Korean, Indian and, especially, Chinese origin started to increase around the same time. On the one hand, this reflects the strong growth of China, India and South Korea which are the main countries of origin of these ethnic groups. Figure \ref{domestic_abroad_shares} in the Appendix highlights this by showing sharply increasing shares of Chinese, Indian and Korean inventors located in their respective countries of origin. On the other hand, it is also well documented in the literature that high-income economies--and particularly the USA--have witnessed large inflows of Asian origin inventors \cite[e.g.][]{miguelez2018, MiguelezFink2013, kerr2007}. Figure \ref{domestic_abroad_absolute}  in the Appendix illustrates this latter channel and shows that especially the number of Chinese and Indian inventors located abroad has massively expanded since the 1990s. Hence, a substantial fraction of the global prevalence increase of these  origin groups is likely due to a surge of Asian origin inventors in high-income countries. At the same time, this has likely contributed to a more diverse inventor population in the respective receiving countries. 
Figure \ref{plot_domestic_share} in the Appenix investigates this in more detail. It plots the evolution of the most dominant ethnic origin, $\hat{\pi}^{k}_{j,t}$, for $j$ corresponding to the six high-income economies USA, Great Britain, Germany, France, Italy and Japan.\footnote{The dominant ethnic origin $k$ corresponds to "Anglo-Saxon" origin for the USA and Great Britain, "German" origin for Germany, "French" origin for France, "Japanese" origin for Japan and "Italian" origin for Italy.} In the $1980$s, the prevalence of the dominant origin was relatively high in most of these countries. The exemption is the USA, which is not surprising as it is an immigration country, traditionally attracting many foreign workers. More interesting differences can be observed over time. The dominant origin share among inventors stayed roughly constant in Japan and Italy, but it decreased substantially for the two Anglo-Saxon countries, Great Britain ($-25$ percentage points) and the USA ($-18$ percentage points), as well as for the two largest European economies, Germany ($-16$ percentage points) and France ($-12$ percentage points).

\begin{figure}[ht]
    \caption{Aggregate Prevalence of Non-Western Ethnic Origins (1980-2015)}
    \centering
    \includegraphics[scale=0.8]{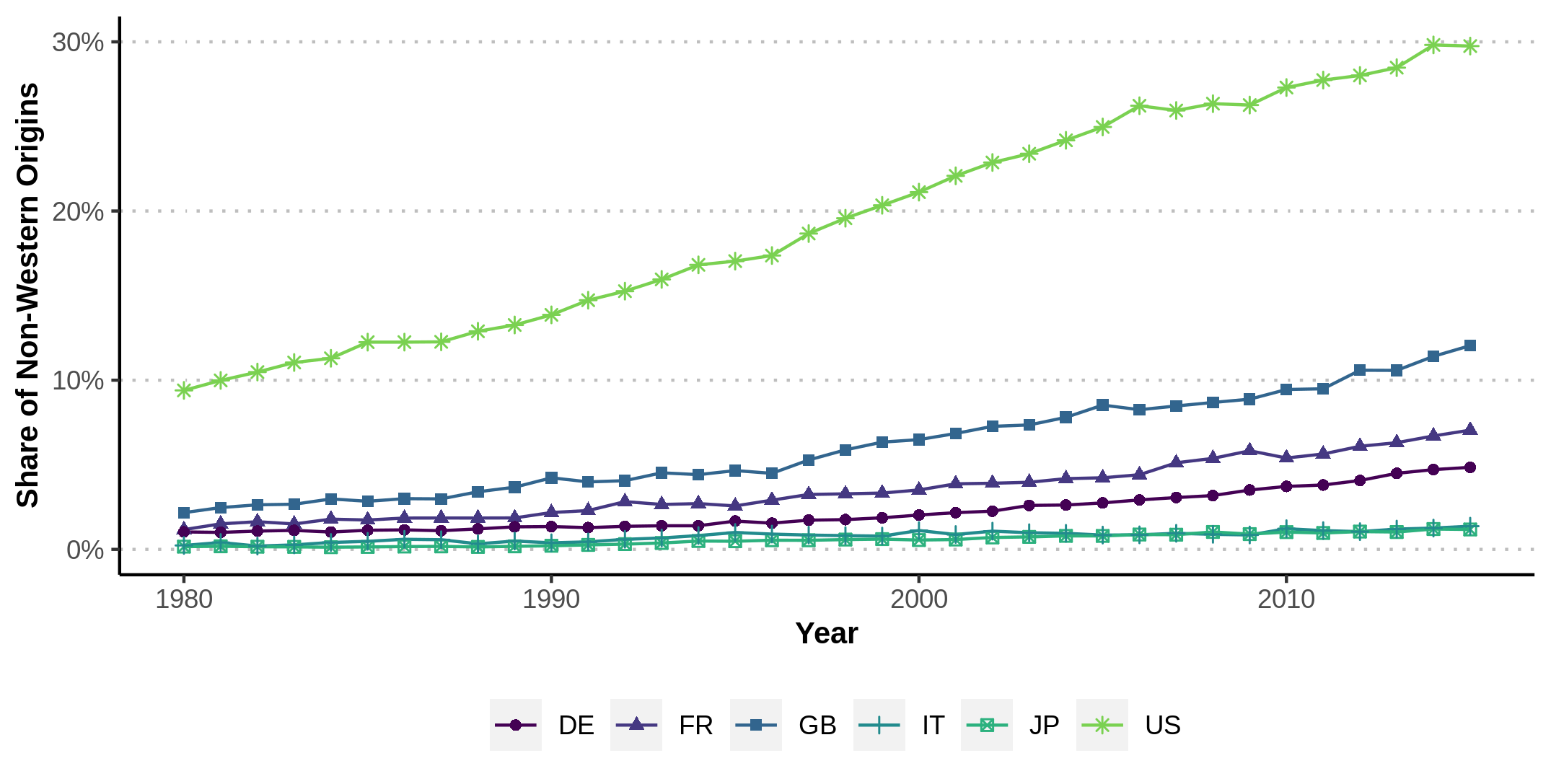}
    \captionsetup{textfont=it}
	\captionsetup{singlelinecheck = false, format= hang, justification=raggedright, font=footnotesize, labelsep=space}
	\caption*{Notes: The graph shows the evolution of the prevalence of non-western ethnic origins among the annual stock of patent inventors in six high-income economies. Formally, it plots $\sum_{k = 1}^{K}\hat{\pi}^{k}_{j,t}$, for each country $j$ with $k$ corresponding to the following non-western ethnic origins: Arabic, Chinese, Indian, Persian, Slavic-Russian, Turkish and South-East Asian. The data for this plot is from the OECD and the USPTO.}
    \label{nonwstern_origins_combined_plot}
    \vspace{-0.5cm}
\end{figure}

Interestingly, this relative decline of the dominant ethnic origin was not driven by an increase of the same foreign origins for the USA and the European countries. Figure \ref{nonwstern_origins_combined_plot} plots the prevalence of non-western ethnic origins in the same six countries.\footnote{The definition of non-western origins is motivated historically, namely that their countries of origin have not been part of the traditional western block of countries during the Cold War and have not joined the European Union in later years. Non-Western ethnic origins are thus defined as Arabic, Chinese, Indian, Persian, Slavic-Russian, Turkish and South-East Asian origin.} Although non-western ethnic origins became more prevalent among patent inventors in all these countries, their share has risen much more in the USA (and to a lesser extent also in Great Britain). In fact, the prevalence of non-western ethnic origins has increased by $20$ percentage points in the USA, which roughly corresponds to the overall decline of the Anglo-Saxon origin. This is different for Great Britain, where the Anglo-Saxon origin has dropped by $25$ percentage points but the share of non-western origins has "only" increased by $10$ percentage points. Even more so for Germany and France. In these countries, the prevalence of non-western origins only increased by around $4$ to $6$ percentage points. Taken together, this suggests that the decline of the dominant ethnic origin was mainly driven by changes within western ethnic origins for European countries, whereas it was induced by inflows of non-western origins for the USA.\footnote{Figure \ref{dominant_origins_European} and \ref{nonwestern_origins_European} in the Appendix illustrate the prevalence of the dominant origin and of non-western origins in other European countries.}

Figure \ref{nonwstern_origins_plot} in the Appendix allows to investigate this in more detail. It breaks down the aggregate prevalence of non-western origin into it's seven ethnic origin components. What can be seen immediately is the well documented inflow of inventors from India and China to the USA starting in the 1990s \cite[see e.g.][]{Peri2016, Kerr2013}. Ethnic origin shares for Chinese and Indian origin have more than tripled in the USA. At the same time, other non-western origins have also increased. In fact, the increase of Indian and Chinese origins accounts for roughly three quarters of the total increase in non-western origin prevalence in the USA. However, this also suggests that the remaining quarter, i.e. increases in other non-western ethnic origins, should not be neglected. In France, half of the total increase in non-western origins was fueled by an inflow of inventors with Arabic roots. In 1980 France, this ethnic group's prevalence was only $0.2$\% compared to $3.3$\% in 2015. Similarly, an inflow of inventors with Indian origin seems to be the main driver for the non-western origin increase in Great Britain. Indian origin corresponded to $1.1$\% in $1980$ and almost quadrupled to $4.2$\% in 2015. In Germany, the picture is more mixed. The largest increase here has come from Slavic-Russian inventors, whose prevalence increased from $0.2$\% to $1.3$\%. 

\begin{figure}[ht]
    \caption{Prevalence of Non-Western Ethnic Origin Across Technologies (1980-2015)}
    \centering
    \includegraphics[scale=0.8]{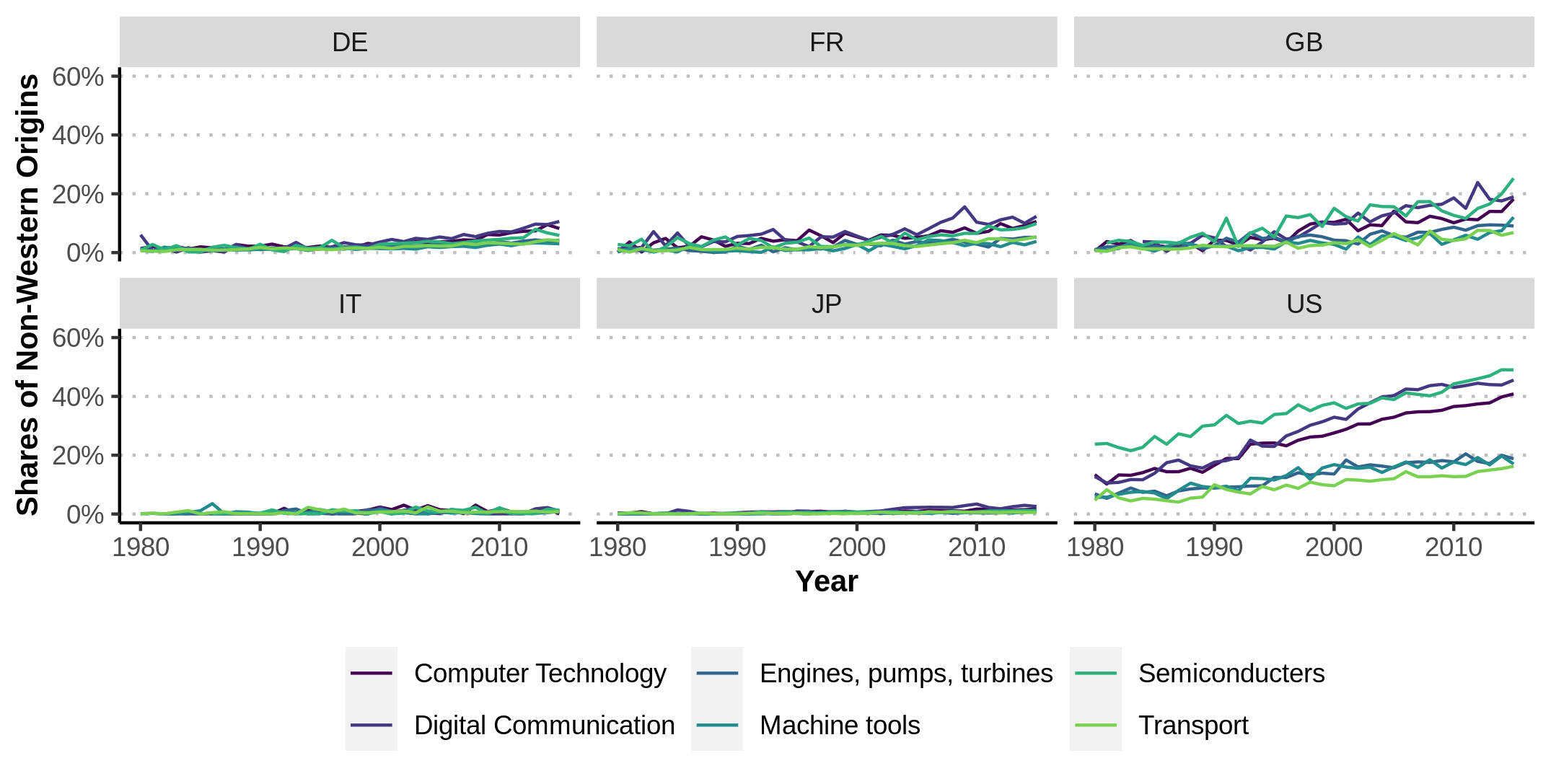}
    \captionsetup{textfont=it}
	\captionsetup{singlelinecheck = false, format= hang, justification=raggedright, font=footnotesize, labelsep=space}
	\caption*{Notes: The graph shows the evolution of the aggregate prevalence of non-western ethnic origins among the annual stock of patent inventors in six technology fields for six high-income economies. Formally, it plots $\sum_{k = 1}^{K}\hat{\pi}^{k}_{j,t}$, for each technology field $j$, with $k$ corresponding to the following non-western ethnic origins: Arabic, Chinese, Indian, Persian, Slavic-Russian, Turkish and South-East Asian. The data for this plot is from the OECD and the USPTO.}
    \label{nonwstern_techfields_US}
    \vspace{-0.5cm}
\end{figure}

Another interesting aspect are differences across technological fields. Figure \ref{nonwstern_techfields_US} states the non-western origin prevalence across six technology fields taken from \cite{schmoch2008}. The graph shows some of the most dynamic technological fields of the recent decades (Computer Technology, Digital Communication, Semiconductors) and more traditional ones (Engines Pumps \& Turbines, Machine Tools, Transport). For the USA, it can immediately be seen that non-western origin inventors are much more prevalent in emerging high-technology fields. This is not only due to initial levels, but because starting around the $1990$s, the non-western origin prevalence has more strongly increased in these technological fields. For example, the prevalence of non-western inventors has increased by more than $25$ percentage points for "Computer Technology", "Semiconductors" or "Digital Communication", reaching a level of over $40$\%. In contrast, the change for "Machine tools" or "Transport" was only about $11$-$12$ percentage points. Looking at European countries, these same patterns cannot be observed to the same extent. Generally, emerging technology fields in these countries do not have a substantially higher prevalence of non-western inventors compared to more traditional ones. What this means is that the USA did not only attract more non-western inventors overall, this inflow was also more directed towards emerging high-technology fields. Several contributions in the economics literature suggest a positive relationship between such high-skilled immigrants and domestic innovation capabilities.\footnote{See, for example, \cite{beerli2021, burchardi2020, akcigit2017,kemeny2017, Kerr2016, Peri2016, Nathan2014, Moser2014,Kerr2013} or \cite{hunt2010} for some recent contributions.} Hence, there is an argument to expect that the larger documented inflows of non-western ethnic origin inventors to the USA have contributed to the country's innovation capabilities and its relative attractiveness for research and development (R\&D) activities, especially in emerging high-technology fields.

\section{Conclusion}\label{conclusion}

The ethnic origin composition of inventors has become more diverse globally and in most high-income economies. But it shows substantial differences across countries and technological fields. In this paper, I have developed a new approach to classify inventors' ethnic origins to study these patterns in detail. 

My proposed classification approach uses inventors' names for ethnic origin classification and builds on a publicly available dataset of athletes who participated in Olympic games between $1896$ and $2016$. Different to the name-matching methods that are more common in the economics literature, my approach is based on supervised learning. In particular, I construct a dataset of $95'202$ labeled names to train an artificial recurrent neural network with long-short-term memory (LSTM), which learns over $1.98$ million parameters and achieves a relatively high performance of $91.0$\% across $17$ ethnic origins when evaluated on a testing sample. 
The proposed classifier addresses several challenges mentioned by previous contributions and seems particularly well suited to tackle the high precision/low recall problem that is inherent to preceding inventor classification methods in the economics literature. A further strength is that it enables to consider a relatively granular level of $17$ ethnic origin groups. This broader origin taxonomy allows me to extend previous analyses from the literature and enables me to study the ethnic origin composition of inventors in detail across countries, technological fields and over a long period of time.

To demonstrate this, I use the trained LSTM network to predict and investigate the ethnic origins of $2.68$ million inventors stated on patents filed at the EPO or the USPTO between $1980$ and $2015$. My descriptive results highlight notable differences over time and across countries and technology fields. The global ethnic origin composition of inventors has become more diverse over the last decades. Especially the prevalence of Chinese, Indian and Korean origin inventors has strongly increased since the $1990$s. In contrast, the prevalence of previously dominant ethnic origins, such as Anglo-Saxon, German or Japanese, has declined. An important part of these changes is due to high-income economies' inventor compositions becoming more ethnically diverse. However, these patterns differ strongly across high-income countries. While some European economies have most likely become more diverse due to migration within Europe, the USA has attracted large inflows of non-western ethnic origin inventors--particularly to emerging high-technology fields. Potentially, this high attractiveness in the global race for talents \cite[e.g.][]{Kerr2016, Kerr2013} is an important advantage for the USA and could have contributed to the country's innovation capacity \cite[e.g.][]{Moser2014, hunt2010}. 

Taken together, there are two particular conclusions that can be drawn from this paper. First, a neural network with LSTM architecture is well suited for classifying inventors' ethnic origins. With this regard, I hope this paper to be an inspiration for future research aiming to determine inventors' ethnic origins in order to investigate, for example, the relationship between migration and innovation. Second, my descriptive analyses regarding the ethnic composition of inventors reveal striking differences across countries and technological fields. Previous research has frequently documented well that high-skilled immigrants foster domestic innovation in the receiving countries \cite[see e.g.][]{bahar2020, cristelli2020, kerrlincoln2010}. Thus, technological clusters which attract higher numbers of non-western patent inventors could have substantially benefited thereof. A related but less explored aspect refers to potential implications in light of the growing internationalization of R\&D activities \cite[see e.g.][]{harhoff2014, griffith2006, bloom2001}. As companies consider the availability of talents for their decisions on where to locate their R\&D activities, differing abilities to attract foreign talents could also affect locations' relative attractiveness for R\&D activities from foreign firms \cite[see e.g.][]{farndale2020,lewin2009, Manning2008}. It is beyond the scope of this paper to provide an in-depth analysis on this particular issue. But I consider it an important and interesting avenue for future research to rigorously investigate the potential interplay between overall migration patterns, inventor mobility and the relative attractiveness of technological clusters for foreign companies' R\&D activities.

\newpage

\section*{Computational Details}
Most computations were performed at sciCORE scientific computing center at the University of Basel (\url{http://scicore.unibas.ch/}). All supervised learning models have been trained and evaluated using \textbf{Python} (Version 3.7) and its libraries \textbf{tensorflow} 2.2 \cite[]{tensorflow_abadi2016}, \textbf{keras} \cite[]{keras_chollet2015}, \textbf{scikit-learn} \cite[]{scikit-learn}, \textbf{pandas} \cite[]{Pandas_mckinney2010} and \textbf{numpy} \cite[]{Numpy_oliphant2006}. 

Data processing and visualisations were mostly conducted in \textbf{R} (Version 4.0.1), whereas I have mainly relied on the package collection \textbf{tidyverse} \cite[]{Tidyverse_wickham2019} and on packages \textbf{data.table} \cite[]{DataTable_Dowle2019} and \textbf{stringi} \cite[]{Stringi_Gagolewski2020} for data processing. Furthermore, I have used the packages \textbf{httr} \cite[]{HTTR_Wickham2019} for accessing the NamePrism-API. Code for using the LSTM model, reproducing the descriptive results as well as example data used for the analysis in this paper is available on GitHub (\url{https://github.com/MatthNig/replication_inventors_ethnic_origins}).

\bibliographystyle{elsarticle-harv}
\bibliography{main}

\newpage

\appendix

\section*{Appendix}

\renewcommand\thefigure{\thesection.\arabic{figure}}
\renewcommand\thetable{\thesection.\arabic{table}}

\section{Constructing Training Samples Using NamePrism\label{TrainingData}}
\setcounter{figure}{0}
\setcounter{table}{0}

The following section provides detailed information on how the $63'585$ sampled inventor names were labeled to construct additional training samples. A very simple approach in this regard would be to build on the taxonomy presented by \cite{Ye2017} and to manually define a crosswalk between the $17$ ethnic origins, $G_k$, used in this paper and the $39$ leaf nationalities from \cite{Ye2017}. Such a crosswalk candidate is presented in Table \ref{NamePrismCrosswalk} in Appendix \ref{TableAppenidx}. With this crosswalk, one could classify a name's ethnic origin according to it's highest NamePrism leaf nationality prediction. Another possibility is to sum up all leaf nationality predictions that are mapped to a particular ethnic origin and to classify the name's ethnic origin according to the highest grouped probability. However, the problem with both of these approaches is that they depend on correctly mapping NamePrism leaf nationalities to the $17$ ethnic origins, $G_k$. Hence, their classification performance should be compared to alternatives. A promising approach are supervised learning algorithms that learn a mapping function between the two taxonomies based on training data.

In order to train such algorithms and to compare their performance to the manual crosswalk approaches, I construct a stratified sample of $12'526$ labeled names from the Olympics dataset and retrieve NamePrism leaf nationality predictions for them. The sample contains $750$ names for all the $17$ ethnic origins expect for Persian origin. For this origin class there are only $526$ names in the athletes dataset, which I have all sampled. Next, I randomly split this data into a training sample of size $10'020$ (i.e. $80\%$ of samples) and a test sample of size $2'506$ (i.e. $20\%$ of samples). I then train three different classifiers on the training set, using the retrieved $39$ leaf nationality predictions from NamePrism as features and evaluate the algorithms' performances on the test set. The three classifiers are a Random Forest, Gradient Boosted Trees (XGBoost) and a Feed-Forward Artificial Neural Network. I have trained these algorithms using Python 3.7 and its libraries scikit-learn, xgboost and tensforflow 2.2. Further, I have used random grid searches and $3$-fold cross-validation to select hyperparameters for every algorithm. Table \ref{MappingAccuracy} shows the tuned algorithms accuracy and F1 score on the test sample and compares them to the two manual crosswalk approaches.

\begin{table}[!h]
\centering
\caption{Mapping Performance: Manual Crosswalks vs. Supervised Learning}
\label{MappingAccuracy}
\begin{threeparttable}
\begin{tabular}{lrr}
\hline\hline
Method & Accuracy & F1 Score\\
\hline
\textit{Manual Crosswalks} & & \\
Highest NamePrism Predicition & $0.690$ & $0.677$\\
Highest Grouped NamePrism Predicition & $0.663$ & $0.653$\\ 
& & \\
\textit{Supervised Learning Algorithms} & & \\
Random Forest & $0.849$ & $0.849$ \\
Gradient Boosted Trees (XGBoost) & $0.852$ & $0.851$ \\
Feed-Forward Neural Network & $0.828$ & $0.827$\\
\hline\hline
\end{tabular}
\begin{tablenotes}
\item \footnotesize{\textit{Notes}: Manually defined crosswalks are presented in Table \ref{NamePrismCrosswalk} in Appendix \ref{TableAppenidx}. Supervised learning algorithms were trained on a set of $10'020$ samples. All approaches were evaluated on the same test set consisting of $2'506$ samples.}
\end{tablenotes}
\end{threeparttable}
\end{table}

Table \ref{MappingAccuracy} clearly highlights supervised learning techniques' power in classification tasks. Whereas the two approaches that are based on manually constructed crosswalks reach an accuracy of $69.0\%$ and $66.3\%$ respectively, machine learning algorithms achieve an accuracy between $80\%$ to $85\%$ on the same test sample. That is to say, they outperform manual classification methods by up to $20$ percentage points. This is because these models are capable to learn much more complex assignment rules that can be highly non-linear. This is something that a manual crosswalk can never achieve. The best performing technique was Gradient Boosted Trees (XGBoost), which has reached a slightly superior performance compared to the Random Forest. Therefore, I will use this model to map NamePrism predictions to the $17$ ethnic origin classes.

However, even though it has a clearly superior performance compared to manual crosswalks, this algorithm still classifies around $15\%$ of samples incorrectly. Recall that these samples will be used to train the final model that classifies ethnic origins based on name patterns. Hence, including all of these samples regardless the expected $15$\% error rate means that the classification model could learn from training data that contains many incorrectly labeled names. Accordingly, the benefit of increasing the amount of training data the model can learn from could be offset because it learns from wrongly labeled data. This could be problematic and would hamper the model's performance. To reduce this concern, only samples that are very clearly assigned to ethnic origins should be added to the athletes dataset and used for learning. To filter to names that the XGBoost has very clearly classified, I construct three metrics for every predicted name: (i) $P_h$, the highest class probability across all $17$ ethnic origin, (ii) $\Delta$, the difference between the highest ($P_h$) and the second highest ethnic origin probability, and (iii) $E$, the entropy across the $17$ ethnic origins.

\begin{table}[!h]
\centering
\caption{Threshold Conditions for Subsetting}
\label{thresholdvalues}
\begin{threeparttable}
\begin{tabular}{lr}
\hline\hline
Metric & Threshold Values\\
\hline
Minimum Value of $P_h$ & \\
(\textit{Highest class probability}) & \{None, 0.45, 0.5, ..., 0.7\}\\
& \\
Minimum Value of $\Delta$ & \\
(\textit{Difference of $P_h$ to 2nd highest origin probability})  & \{None, 0.2, 0.3, ..., 0.5\} \\
& \\
Maximum Value of $E$ & \\
(\textit{Entropy across $17$ ethnic origins}) & \{1.75, 2, None\}\\
\hline\hline
\end{tabular}
\begin{tablenotes}
\item \footnotesize{\textit{Notes}: Overview of the different threshold conditions for subsetting the data to very clearly classified inventor names. Together, these threshold conditions define a set of $7 \times 5 \times 3 = 105$ combinations.} 
\end{tablenotes}
\end{threeparttable}
\end{table}

Together, these metrics indicate if a name has been relatively unambiguously classified by the XGBoost algorithm. I can discard all samples that do not fulfill certain threshold conditions based on these metrics. Naturally, there is a trade-off between applying strict conditions, which leads to discarding a larger proportion of samples (false negatives) and the possibility to include many wrongly labeled names (false positives). A solution to this can be best approximated empirically and I proceed in the following way: For each of the three metrics I define a range of threshold values which are presented in Table \ref{thresholdvalues}. This provides $105$ different threshold combinations (i.e., $7 \times 5 \times 3 = 105$), which I can use to subset the data.

For every threshold combination, the resulting subset has a different size, where all are smaller than the original test sample of $2'506$ names. The goal is then to find a threshold combination that creates a sample with a relatively high classification performance, but only drops relatively few samples. To approximate both aspects in every of the $105$ sub-samples, I calculate it's overall F1 score and three different indicators capturing the impact of subsetting on the sample size and sample structure. The latter three indicators are the sample fraction relative to the baseline test sample, the variance of the sample shares of the ethnic origin groups and the combined sample share of the two smallest ethnic origin classes. The reason to consider the latter two as evaluation metrics is that it could distort the performance of the final model if a threshold combination would completely drop smaller ethnic origin classes in return for a higher performance. For comparison, I standardize each of the four obtained metrics on a zero-one scale and weight them to compute a score for every of the $105$ threshold combinations.\footnote{I weight the four metrics as follows: $50$\% for the F1 score, $25$\% for the sample fraction relative to the baseline and 12.5\% each for the two metrics considering the origin shares. To test the robustness of the evaluation result, I have also varied these weights using $26$ different weighting schemes to compute the evaluation score. It is reassuring that across the $105$ combination candidates, my optimally chosen threshold combination is ranked among the $5$ best performing combinations in $50\%$ of all these $26$ alternative weighting schemes (in $69\%$ among the $20$ best performing combinations).}

\begin{figure}[!h]
    \caption{Evaluation of Threshold Combinations}
    \centering
    \includegraphics[scale=0.8]{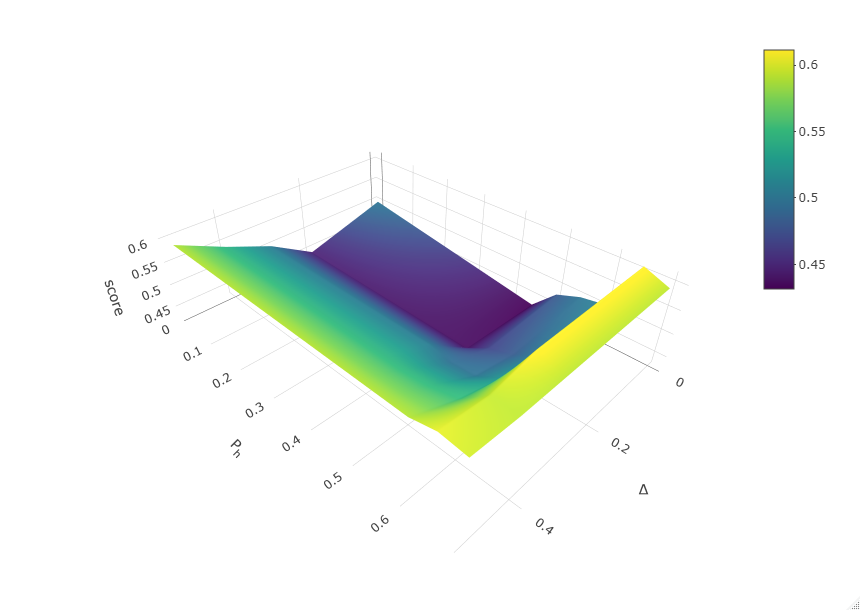}
    \captionsetup{textfont=it}
	\captionsetup{singlelinecheck = false, format= hang, justification=raggedright, font=footnotesize, labelsep=space}
	\caption*{Notes: The graph's color and z-axis show the performance scores of $105$ different threshold combinations. $P_h$ depicts threshold values for the highest class probability estimated by the XGBoost algorithm, $\Delta$ states the difference from the first to the second highest class probability.}
    \label{plot_threshold_eval}
\end{figure}

Figure \ref{plot_threshold_eval} plots these scores for all the $105$ threshold combinations. The highest score was obtained by subsetting the data based on the following threshold values: The highest class probability across all $17$ ethnic origin, $P_h$ must be equal or higher than $0.65$, the difference between the highest and the second highest ethnic origin probability, $\Delta$, must be equal or larger than $0.2$ and no additional restrictions are introduced on the value of the entropy across the $17$ ethnic origins. Filtering the data according to this combination yields an F1-score of $0.927$ on the resulting sample, which retains $84.5$\% of the original test sample's size. In other words, applying the XGBoost algorithm to new data \textit{and} only retaining those samples that fulfill $P_h \ge 0.65$ and $\Delta \ge 0.2$ is expected to result in an overall classification performance of $0.927$ (instead of $0.851$), while reducing the number of samples to $84.5$\% of the original size. The final step to complete the construction of the training data is to exactly do this for the sampled $63'585$ inventor names, which I have NamePrism leaf nationality information. First, I use the trained XGBoost algorithm to predict the ethnic origin of each inventor name based on the $39$ NamePrism leaf nationality predictions. Second, I apply the best performing threshold combination and drop all samples that do not fulfill $P_h \ge 0.65$ or $\Delta \ge 0.2$. This results in a dataset of $53'189$ names, which is $83.6$\% of the size of the original $63'585$ inventor names sample (i.e. it is close to the expected fraction of $84.5$\%). Ultimately, I combine these inventor names with the $42'013$ labeled athletes names. Hence, my final dataset consists of $95'202$ samples across $17$ ethnic origins.

\newpage

\section{Origin Classification Using a LSTM \label{LSTMModel}}
\setcounter{figure}{0}

This section documents the classification model and it's training process. The basis for the latter are the $95'202$ labeled names described in Section \ref{classification_sec} and Appendix \ref{TrainingData}. I follow recommendations from the literature \cite[]{Chollet2018} and split this dataset into three samples: In a first step, I select $10$\% of names for testing by separating them from the rest of the data. That is, $9'521$ names are reserved as out of sample data and $85'681$ remain for training. Next, I split the remaining names into a training and a validation set. This step is important for finding a useful model architecture (e.g. the number of LSTM layers) and choosing the network's hyperparameters (e.g. the learning rate of the optimizer). Networks with different architectures and hyperparameters are trained on the training set and their classification performances are evaluated and compared based on the validation set \cite[see e.g.][for the importance of using a validation set]{Chollet2018}. I assign $85$\% of the $85'681$ remaining names for training ($72'828$) and $15$\% for validation ($12'853$).

I have used Python and its deep learning libraries Keras \cite[]{keras_chollet2015} and Tensorflow 2.2 \cite[]{tensorflow_abadi2016} to train different networks.\footnote{Models were trained on GPUs either at Google Colaboratory or at sciCORE scientific computing center at the University of Basel.} After experimenting with several specifications and hyperparamerters, I chose a network consisting of $3$ LSTM-layers containing $512$, $264$ and $64$ LSTM-cells, respectively. All LSTM-cells feature a dropout of $0.2$ to counter overfitting. On top of these $3$ LSTM-layers, I add an output layer with softmax activation that relates the inputs to the $17$ ethnic origin classes. In total, this models learns over $1.98$ million parameters. For training, I use sparse categorical crossentropy loss and an adam optimizer with a learning-rate of $0.0025$. Training is conducted in batches of $256$ samples over a maximum of $50$ epochs with early stopping after $7$ epochs without a performance increase. As recommended in the literature \cite[e.g.][]{Chollet2018}, I combine the training and validation set to train the final model and evaluate it's performance on the testing set it has never seen before based on F1-scores. The trained network achieves a weighted F1-score of $0.91$ on the testing set and performs relatively similar across ethnic origin classes (see Section \ref{classification_sec}). In a last step, I follow recommendations from the literature \cite[]{Chollet2018} and train the network on all the $95'202$ available names. Subsequently, I use this fully trained model to predict the ethnic origins of $2.68$ million patent inventors (see Section \ref{ethnic_sec}).

\section{Patent Data\label{PatentData}}

The dataset of patents used in this paper has been constructed at the Center for International Economics and Business $|$ CIEB at the University of Basel. The sample I use from this dataset consists of over $7.7$ million patents filed at the USPTO or the EPO between $1980$ and $2015$.\footnote{All the data is publicly available and can be accessed using Patentsview \cite[]{uspto_data} and the OECD patent database \cite[]{oecd_data}. For constructing the final dataset, patents from the two sources have been cleaned of equivalent patents \cite[see e.g. ][for a detailed discussion]{webb2005}.} For the analyses in this paper, I use the names of over $2.68$ million inventors stated on patents in this dataset and extract their country of residence. Additionally, I gather information on the patents these inventors have filed: I collect the technology field a patent has been assigned to \cite[see][]{schmoch2008} and the year it has been filed first anywhere in the world (the so-called "priority year"). Combining all this information allows me to perform the analyses presented in Sections \ref{ethnic_sec}.

\newpage

\section{Additional Figures}\label{AppendixGraphs}
\setcounter{figure}{0}

\begin{figure}[!h]
    \caption{Example of the Encoded Name Mahatma Gandhi}
    \centering
    \includegraphics[width=14cm,height=9cm]{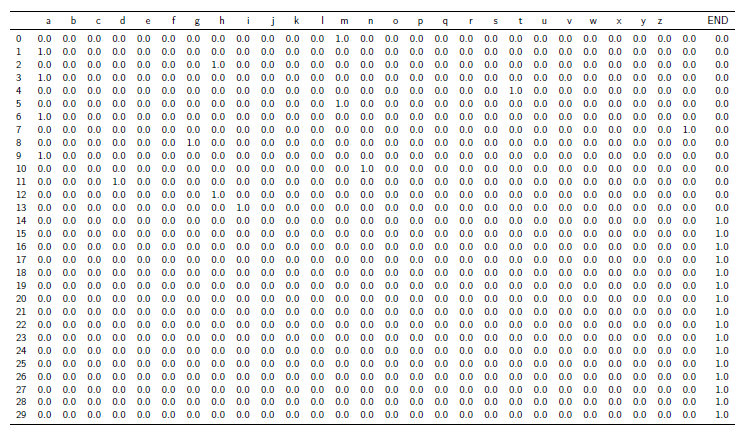}
    \label{gandhi_example}
    \vspace{-0.5cm}
\end{figure}

\begin{figure}[!ht]
    \caption{Visualization of an LSTM cell \cite[taken from][]{Goodfellow2016}}
    \centering
    \includegraphics[scale=0.5]{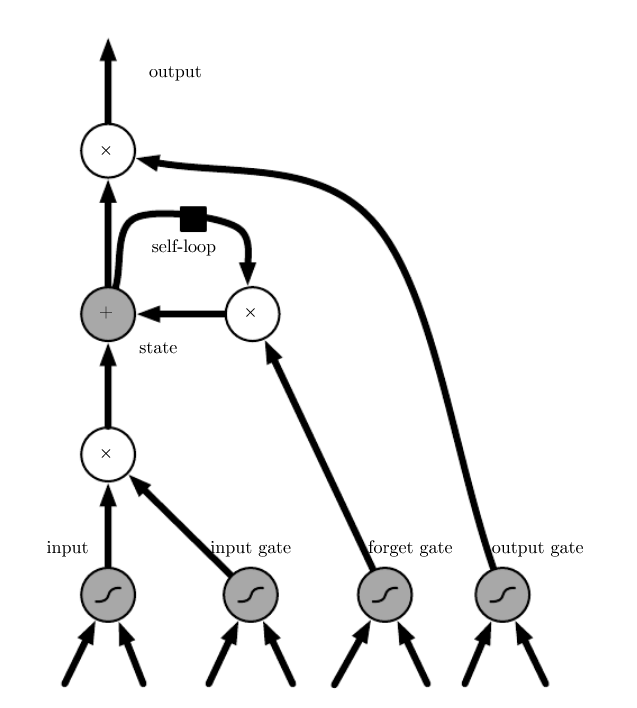}
    \captionsetup{textfont=it}
	\captionsetup{singlelinecheck = false, format= hang, justification=raggedright, font=footnotesize, labelsep=space}
	\caption*{Notes: The graph shows an illustration of an LSTM-cell taken from \cite{Goodfellow2016}. Input, input gate and forget gate jointly determine the cell state. The output gate and the cell state define the LSTM-cell's output at timestep $t$.}
    \label{lstm_cell}
    \vspace{-0.5cm}
\end{figure}

\begin{figure}[ht]
    \caption{Location of Chinese, Indian and Korean Inventors (1980-2015)}
    \centering
    \includegraphics[scale=0.8]{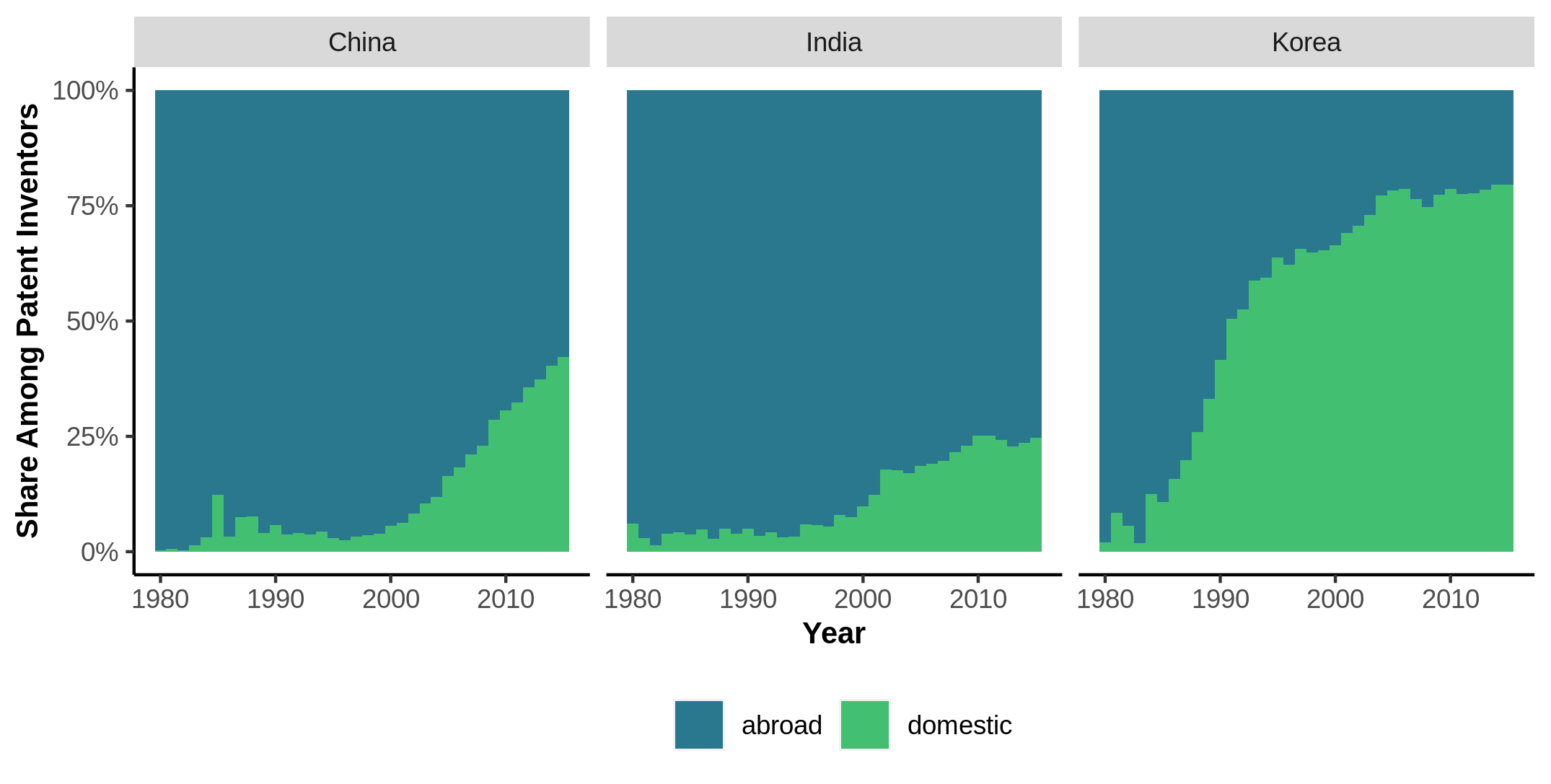}
    \captionsetup{textfont=it}
	\captionsetup{singlelinecheck = false, format= hang, justification=raggedright, font=footnotesize, labelsep=space}
	\caption*{Notes: The graph shows the annual estimated share of Chinese, Indian and Korean patent inventors located domestically and abroad between 1980 and 2015. The underlying data consists of inventors whose highest ethnic origin prediction obtained from the LSTM model, $\hat{G}_i$, corresponds to Chinese, Indian or Korean origin. These inventors are then classified as "domestic" if their residence address is located within China, India or South Korea, respectively and as abroad otherwise. The data for this plot is from the OECD and the USPTO.}
    \label{domestic_abroad_shares}
    \vspace{-0.5cm}
\end{figure}

\begin{figure}[ht]
    \caption{Number of Chinese, Indian and Korean Inventors (1980-2015)}
    \centering
    \includegraphics[scale=0.8]{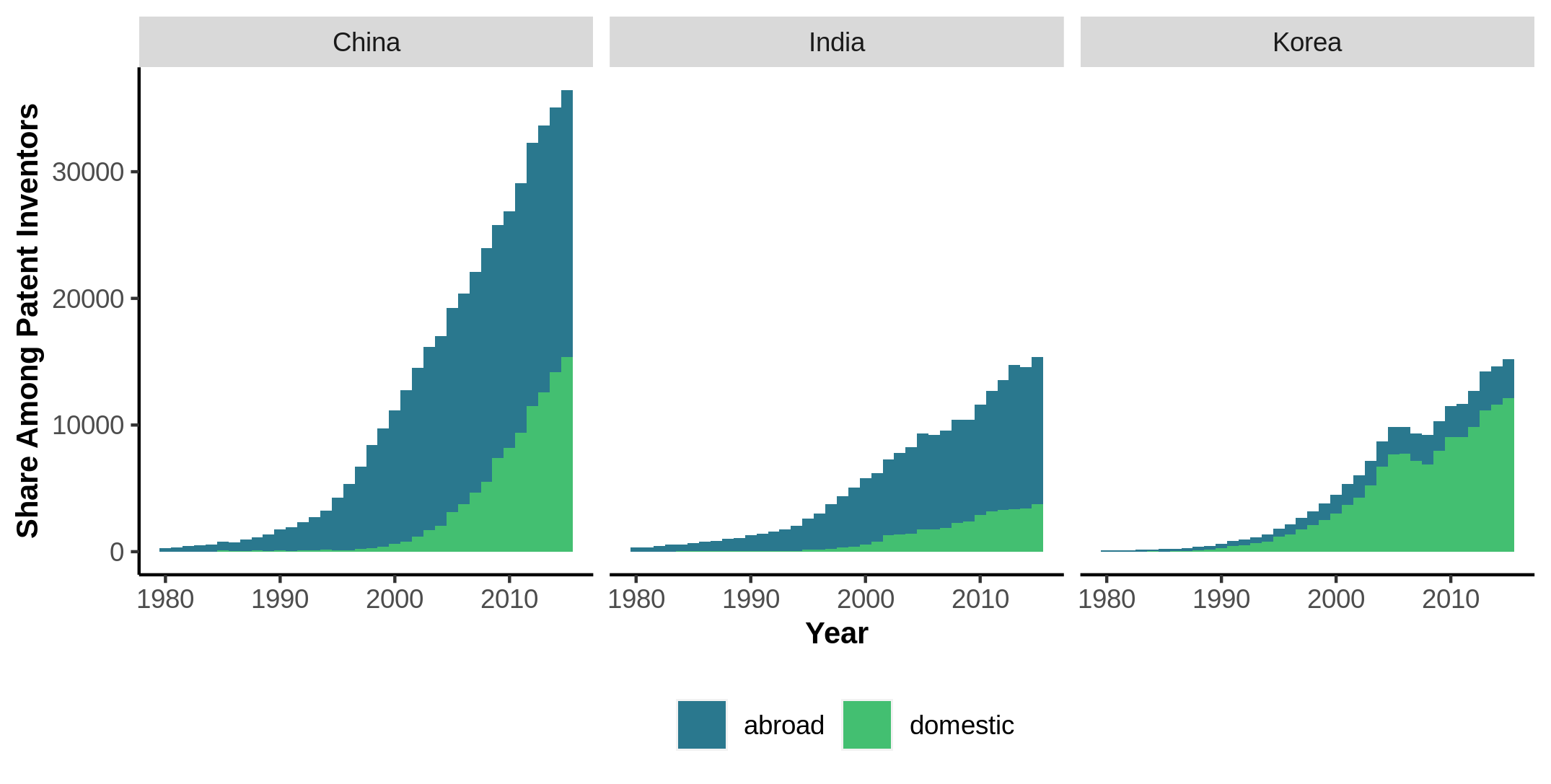}
    \captionsetup{textfont=it}
	\captionsetup{singlelinecheck = false, format= hang, justification=raggedright, font=footnotesize, labelsep=space}
	\caption*{Notes: The graph shows the annual estimated number of Chinese, Indian and Korean patent inventors located domestically and abroad between 1980 and 2015. The underlying data consists of inventors whose highest ethnic origin prediction obtained from the LSTM model, $\hat{G}_i$, corresponds to Chinese, Indian or Korean origin. These inventors are then classified as "domestic" if their residence address is located within China, India or South Korea, respectively and as abroad otherwise. The data for this plot is from the OECD and the USPTO.}
    \label{domestic_abroad_absolute}
    \vspace{-0.5cm}
\end{figure}

\begin{figure}[ht]
    \caption{Prevalence of the Dominant Ethnic Origin (1980-2015)}
    \centering
    \includegraphics[scale=0.8]{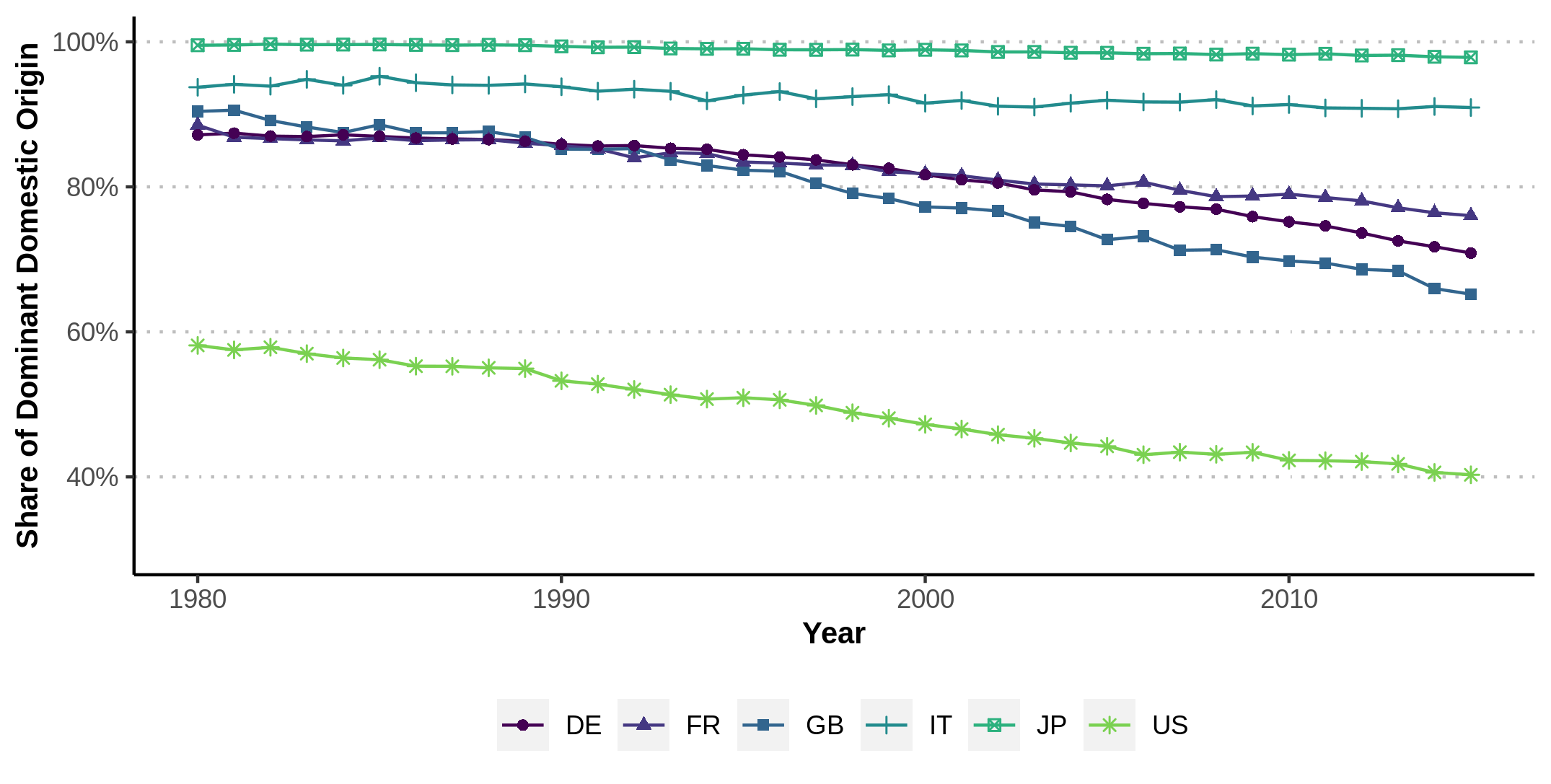}
    \captionsetup{textfont=it}
	\captionsetup{singlelinecheck = false, format= hang, justification=raggedright, font=footnotesize, labelsep=space}
	\caption*{Notes: The graph shows the evolution of the most dominant ethnic origin for six high-income economies. Formally, it plots $\hat{\pi}^{k}_{j,t}$, with $k$ corresponding to Anglo-Saxon for the USA and Great Britain, German for Germany, French for France, Japanese for Japan and Italian for Italy. The data for this plot is from the OECD and the USPTO.}
    \label{plot_domestic_share}
    \vspace{-0.5cm}
\end{figure}

\begin{figure}[ht]
    \caption{Prevalence of Non-Western Ethnic Origins (1980-2015)}
    \centering
    \includegraphics[scale=0.8]{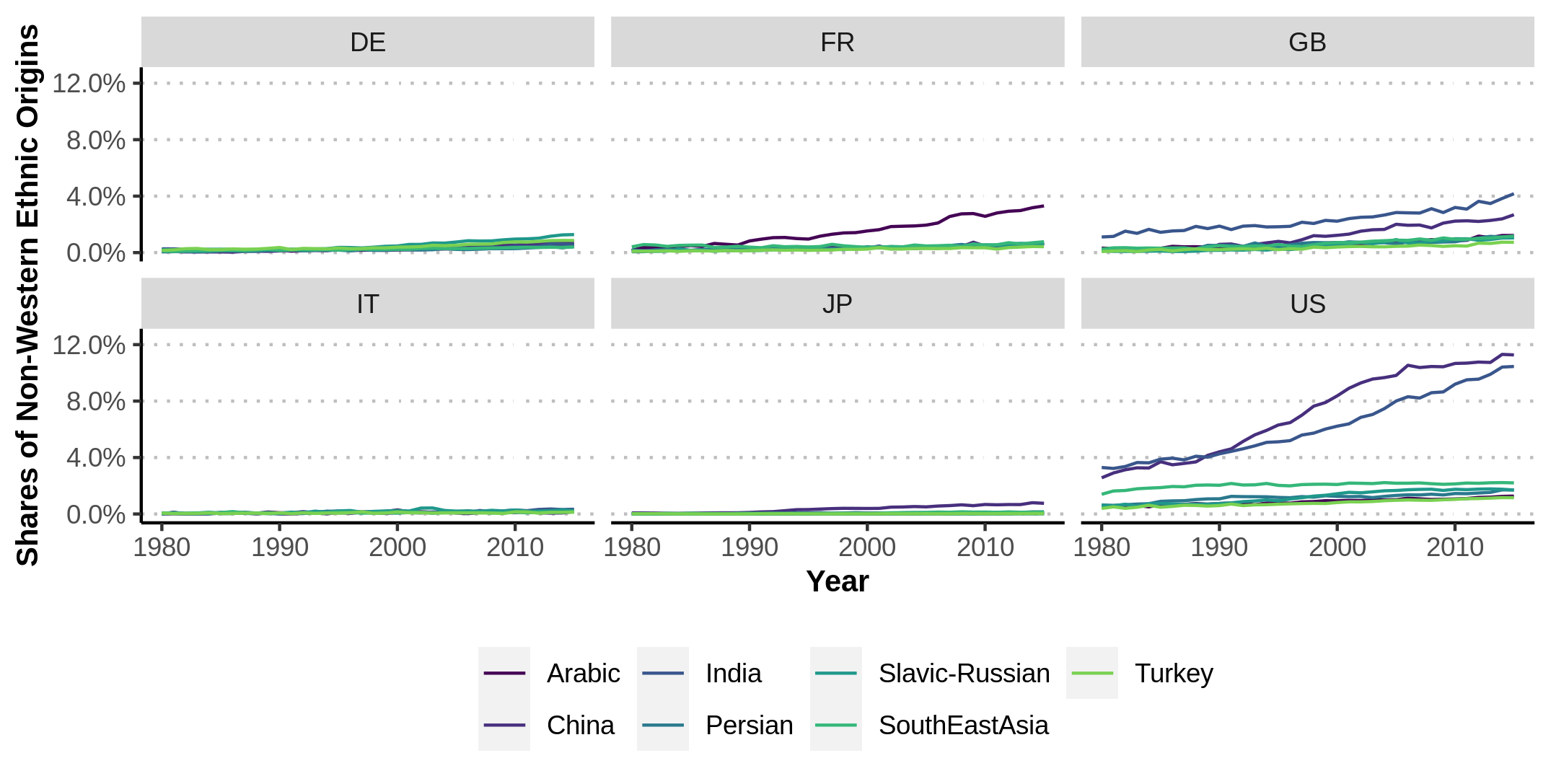}
    \captionsetup{textfont=it}
	\captionsetup{singlelinecheck = false, format= hang, justification=raggedright, font=footnotesize, labelsep=space}
	\caption*{Notes: The graph shows the evolution of the prevalence of non-western ethnic origins in four high-income countries. Formally, it plots $\hat{\pi}^{k}_{j,t}$, for each country $j$, with $k$ corresponding to the following non-western ethnic origins: Arabic, Chinese, Indian, Persian, Slavic-Russian, Turkish and South-East Asian. The data for this plot is from the OECD and the USPTO.}
    \label{nonwstern_origins_plot}
    \vspace{-0.5cm}
\end{figure}

\begin{figure}[!ht]
    \caption{Prevalence of the Dominant Ethnic Origin in European Countries (1980-2015)}
    \centering
    \includegraphics[scale=0.8]{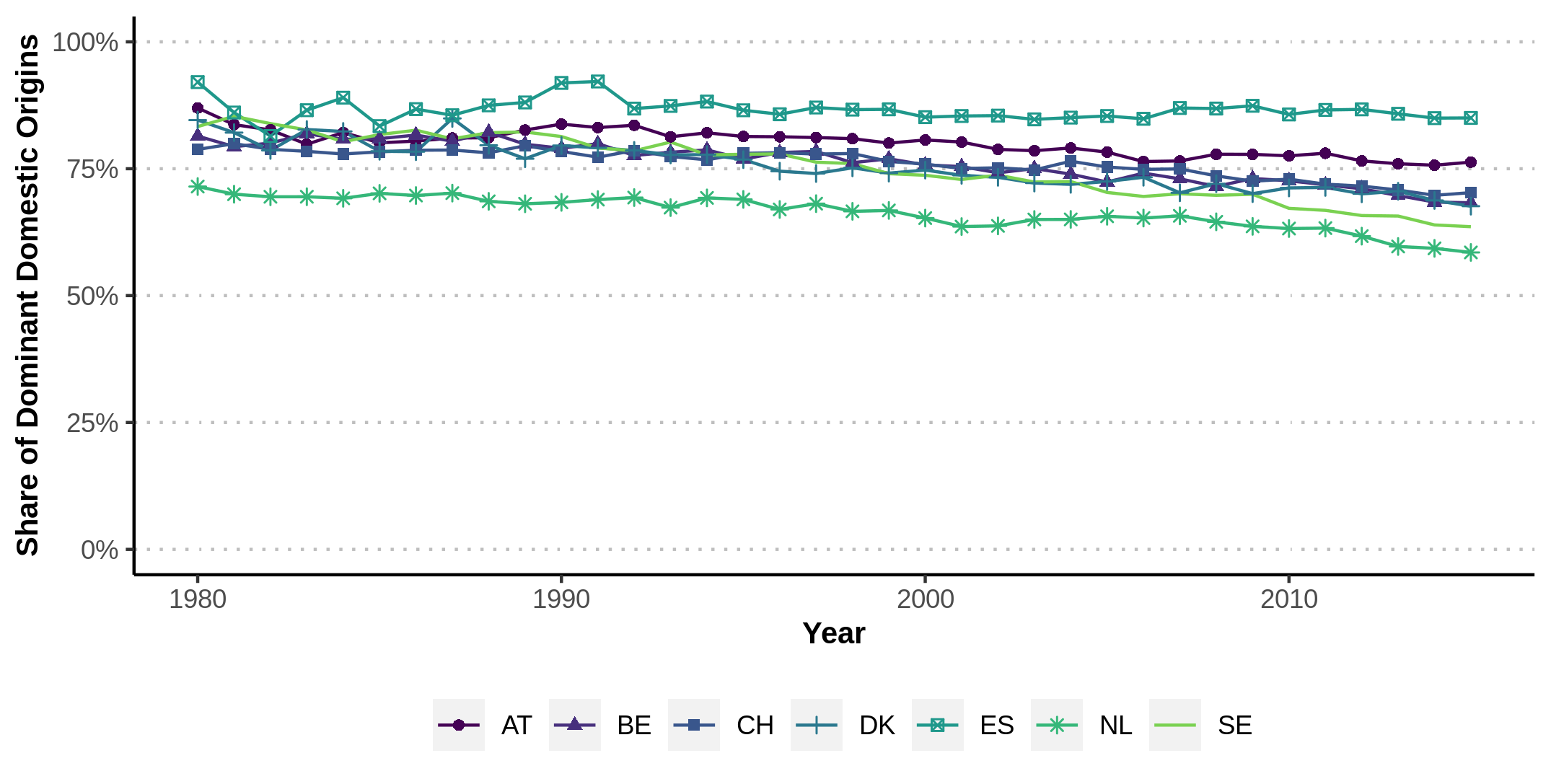}
    \captionsetup{textfont=it}
	\captionsetup{singlelinecheck = false, format= hang, justification=raggedright, font=footnotesize, labelsep=space}
	\caption*{Notes: The graph shows the evolution of the prevalence of the most dominant ethnic origins for seven European countries. Formally, it plots $\pi_{kt}$, with $k$ corresponding to German assigned for Austria, Switzerland and the Netherlands, French for Switzerland and Belgium, Scandinavian for Denmark and Sweden, Anglo-Saxon for the Netherlands. The data for this plot is from the OECD and the USPTO.}
    \label{dominant_origins_European}
    \vspace{-0.5cm}
\end{figure}

\begin{figure}[!ht]
    \caption{Prevalence of Aggregate Non-Western Ethnic Origins in European Countries (1980-2015)}
    \centering
    \includegraphics[scale=0.8]{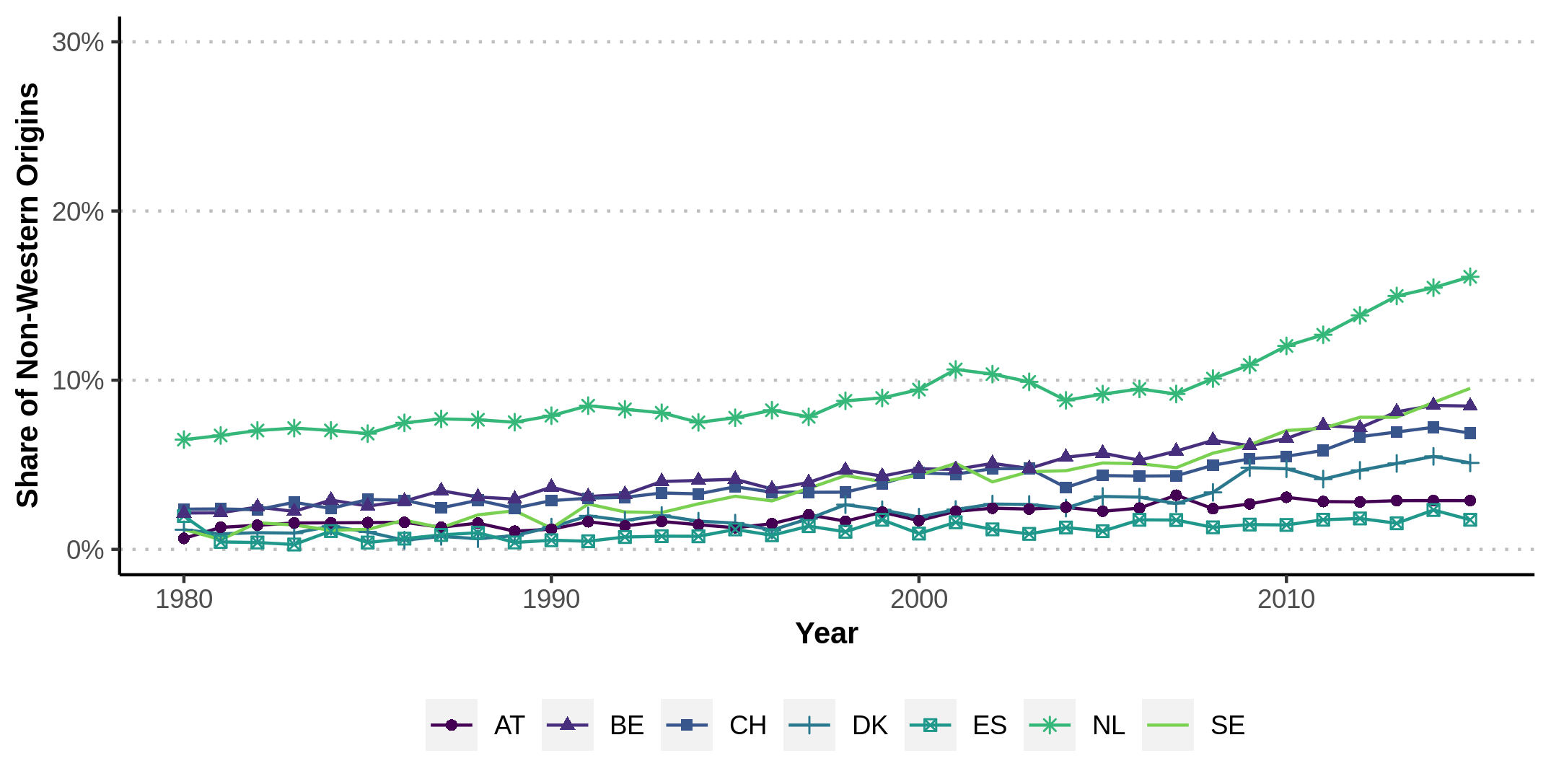}
    \captionsetup{textfont=it}
	\captionsetup{singlelinecheck = false, format= hang, justification=raggedright, font=footnotesize, labelsep=space}
	\caption*{Notes: The graph shows the evolution of the aggregate prevalence of non-western ethnic origins in seven European countries. Formally, it plots $\sum_{k = 1}^{K}\pi_{kt}$, for each country with $k$ corresponding to the following non-western ethnic origins: Arabic, Chinese, Indian, Persian, Slavic-Russian, Turkish and South-East Asian. The data for this plot is from the OECD and the USPTO.}
    \label{nonwestern_origins_European}
    \vspace{-0.5cm}
\end{figure}

\clearpage

\section{Additional Tables}\label{TableAppenidx}
\setcounter{table}{0}

\begin{table}[h]
\centering
\begin{threeparttable}
\caption{Manual Crosswalk: Ethnic Origins and NamePrism Leaf Nationalities}
\label{NamePrismCrosswalk}
\begin{tabular}{ll}
\hline\hline
Ethnic Origin& NamePrism Leaf Nationality \\ 
\hline
&\\
AngloSaxon & Celtic-English \\ 
Arabic & Muslim, Pakistanis, Bangladesh \\ 
Arabic & Muslim, Maghreb \\ 
Arabic & Muslim, Pakistanis, Pakistan \\ 
Arabic & Muslim, ArabianPeninsula \\ 
Balkans & European, SouthSlavs \\ 
China & EastAsian, Chinese \\ 
EastEurope & European, Baltics \\ 
EastEurope & European, EastEuropean \\ 
French & European, French \\ 
German & European, German \\ 
Hispanic-Iberian & Hispanic, Portuguese \\ 
Hispanic-Iberian & Hispanic, Spanish \\ 
India & SouthAsian \\ 
Italian & European, Italian, Italy \\ 
Italian & European, Italian, Romania \\ 
Japan & EastAsian, Japan \\ 
Korea & EastAsian, South Korea \\ 
Persian & Muslim, Persian \\ 
Scandinavian & Nordic, Scandinavian, Denmark \\ 
Scandinavian & Nordic, Finland \\ 
Scandinavian & Nordic, Scandinavian, Sweden \\ 
Scandinavian & Nordic, Scandinavian, Norway \\ 
Slavic & European, Russian \\ 
SouthEastAsia & EastAsian, Indochina, Thailand \\ 
SouthEastAsia & EastAsian, Indochina, Vietnam \\ 
SouthEastAsia & EastAsian, Indochina, Cambodia \\ 
SouthEastAsia & EastAsian, Indochina, Myanmar \\ 
SouthEastAsia & EastAsian, Malay, Malaysia \\ 
SouthEastAsia & EastAsian, Malay, Indonesia \\ 
Turkey & Muslim, Turkic, Turkey \\ 
\hline\hline
\end{tabular}
\begin{tablenotes}[flushleft]
\item \footnotesize{\textit{Notes:} Manually constructed crosswalk between the $17$ ethnic origins in this paper and the $39$ leaf nationalities from NamePrism \cite[]{Ye2017}. For further information, see Section 2 and Appendix \ref{TrainingData}).} 
\end{tablenotes}
\end{threeparttable}
\end{table}

\begin{table}[]
\centering
\begin{threeparttable}
\caption{Ethnic Origin Distribution in the Training Data}
\label{TrainingDataDistribution}
\begin{tabular}{lrr}
\hline\hline
Ethnic Origin & Number of Samples & Sample Fraction \\
\hline
Anglo-Saxon & 7'933 & 0.083 \\
Arabic & 3'795 & 0.040 \\
Balkans & 2'320 & 0.024\\
Chinese & 6'567 & 0.069\\
East-Europe & 6'820 & 0.072\\
French & 7'737 & 0.081\\
German & 6'311 & 0.066\\
Hispanic-Iberian & 6'383 & 0.067\\ 
India & 4'205 & 0.044 \\
Italian & 6'171 & 0.065\\ 
Japanese & 8'835 & 0.093\\ 
Korean & 5'917 & 0.062\\ 
Persian & 1'614 & 0.017\\
Scandinavian & 6'938 & 0.073\\
Slavic-Russian & 6'357 & 0.067 \\
South-East Asia & 2'895 & 0.030\\
Turkish & 4'404 & 0.046\\
\textbf{Total} & \textbf{95'202} & \textbf{1.000} \\
\hline\hline
\end{tabular}
\begin{tablenotes}[flushleft]
\item \footnotesize{\textit{Notes:} The training data consists of $53'189$ (55.9\%) names of athletes from the Olympics dataset described in Section 2 and $42'013$ (44.1\%) inventor names that were labeled using NamePrism (see Appendix \ref{TrainingData}).} 
\end{tablenotes}
\end{threeparttable}
\end{table}

\end{document}